\documentclass{elsart}

\usepackage{natbib}

\usepackage{epsfig}

\usepackage{amssymb}

\def\eps@scaling{.95}
\def\epsscale#1{\gdef\eps@scaling{#1}}
\def\plotone#1{\centering \leavevmode
\epsfxsize=\eps@scaling\columnwidth \epsfbox{#1}}

\begin{document}

\begin{frontmatter}



\title{AGN Host Galaxies}


\author{Sylvain Veilleux}

\address{University of Maryland}

\begin{abstract}
  In this series of four lectures, I discuss four important aspects of
  AGN host galaxies. In Lecture \#1, I address the starburst-AGN
  connection. First, I briefly review the primary diagnostic tools
  that are used to quantify and distinguish star formation and nuclear
  activity.  Next I describe the best evidence for a connection between
  these two processes, first at low luminosity and then at high
  luminosity. In the last section, I summarize the main results and
  offer possible explanations. In Lecture \#2, I discuss our current
  understanding of ultraluminous infrared galaxies
  (log[L$_{IR}$/L$_\odot$] $\ge$ 12; ULIRGs). First, I describe the
  general properties of ULIRGs, comparing the local sample with their
  distant counterparts. Then I discuss the role of ULIRGs in the
  formation and evolution of spheroids and their massive black holes.
  The discussion of their possible role in the metal enrichment of the
  IGM through superwinds is postponed until Lecture \#3. In this third
  lecture, I discuss the importance of feedback processes in the local
  and distant universe.  The emphasis is on mechanical feedback. I
  describe the basic physics of winds, a few classic examples of winds
  in the local universe, the statistical properties of winds, near and
  far, and their impact on galaxy formation and evolution. A list of
  potential thesis projects is given at the end.  The fourth and final
  lecture is on elemental abundances as tracers of star formation.
  First, I explain the basic principles behind chemical evolution, and
  describe three simple models whose predictions are compared with
  observations in the Milky Way. Next I discuss and give an
  interpretation of the results of abundance determinations in local
  quiescent and starburst galaxies before discussing elemental
  abundances in the more distant universe.
\end{abstract}


\end{frontmatter}


\section{The Starburst-AGN Connection}

\subsection{Introduction}
\label{}

The apparent connection between black hole driven nuclear activity and
starburst activity on large scale has been the topic of debates for
many years (e.g., see references in review by Veilleux 2001). More
than ever, this topic is relevant to help us understand galaxy formation
and evolution, the global star formation and metal enrichment history
of the universe, and the origin of nuclear activity and associated
black hole growth.

The existence of an apparently tight relation (Fig. 1) between black
hole masses and spheroid masses (or velocity dispersions; Gebhardt et
al.  2000; Ferrarese \& Merritt 2000) points to a causal connection
between spheroid formation (via a starburst) and black hole growth
(via nuclear activity). A flurry of theoretical papers have tried to
make sense of these results. In many scenarios, gas or radiation
pressure from a starburst- and/or AGN-driven wind helps shut off the
fuel supply to the black hole and terminate star formation in the
surrounding galaxy (e.g., Murray, Quataert, \& Thompson 2005).
Regardless of the exact process involved in regulating the black hole
and spheroid growths (this topic of negative feedback is covered in
Lecture \#3 of this series, \S 3), the correlation indicates that the
starburst-AGN connection is alive and well and has had a
cosmologically important impact on galaxy formation and evolution.

\begin{figure}
\caption{Black hole mass versus bulge luminosity (left) and the
  luminosity-weighted aperture dispersion within the effective radius
  (right). Green squares
  denote galaxies with maser detections, red triangles are from gas
  kinematics, and blue circles are from stellar kinematics. Solid and
  dotted lines are the best-fit correlations and their
  68\% confidence bands. (From Gebhardt et al. 2000)}
  \end{figure}

To better understand this connection, one first needs to discuss the
diagnostic tools that are used to detect and distinguish star
formation and nuclear activity. This is done in \S 1.2. In \S 1.3, I
describe key results from recent studies of low- and high-luminosity
AGNs. In \S 1.4, I summarize the results and offer a few possible
explanations.

\subsection{Star Formation Diagnostics}
\label{}

A very useful paper here is Kennicutt (1998). The material in \S\S 1.2.1
-- 1.2.4 is taken directly from that review and is therefore not
described in detail. 

\subsubsection{Ultraviolet}

Hot, young stars emit copious amounts of UV radiation (e.g., Leitherer
et al. 1999). The strength of the UV (1500 -- 2800 \AA) continuum
scales linearly with the luminosity of young stars and therefore with
the star formation rate. For solar abundances and a Salpeter Initial
Mass Function ($\equiv$ IMF, 0.1 -- 100 M$_\odot$):

\begin{eqnarray*}
  SFR (M_\odot~{\rm yr}^{-1}) = 1.4 \times 10^{-28} L_\nu ({\rm ergs}~{\rm s}^{-1} {\rm Hz}^{-1})
\end{eqnarray*}

\subsubsection{Recombination Lines}

Hot, young stars emit radiation that ionizes the surrounding ISM. The
Str\"omgren sphere is the spherical volume of this HII region where
the rate of ionizations balances the rate of recombinations.  The
nebular lines produced in the HII region effectively re-emit the
integrated stellar luminosity shortward of the Lyman limit ({\em i.e.}
$\ge$ 13.6 eV). The intensity of these lines scales linearly with the
number of hot, young stars and therefore the star formation rate. For
solar abundances and a Salpeter IMF (0.1 -- 100 M$_\odot$):

\begin{eqnarray*}
SFR (M_\odot~{\rm yr}^{-1}) = 7.9 \times 10^{-42} L_{H\alpha} ({\rm
ergs}~{\rm s}^{-1})
\end{eqnarray*}

\subsubsection{Forbidden Lines}

H$\alpha$ is redshifted out of the visible window beyond $z \sim 0.5$.
In principle, H$\beta$ and the higher order Balmer emission lines
could be used to estimate star formation rates, but these lines are
weak and stellar absorption more strongly influences their
emission-line fluxes than that of H$\alpha$.

Neutral oxygen has the same ionization potential as hydrogen (13.6
eV). This means that ionized oxygen coexists with ionized hydrogen and
therefore lines produced by ionized oxygen scales with the number of
hot, young stars in HII regions. The strongest emission feature in the
blue is the [O II] $\lambda\lambda$3726, 3729 forbidden-line
doublet. The strength of these collisionally excited lines is
sensitive to abundance and ionization state (electron temperature) of
the gas, more so than the recombination lines. A rough calibration is:

\begin{eqnarray*}
SFR (M_\odot~{\rm yr}^{-1}) = (1.4 \pm 0.4) \times 10^{-41} L_{[O II]}
({\rm ergs}~{\rm s}^{-1})
\end{eqnarray*}

Given its blue wavelength, this doublet is also more sensitive to dust
extinction than H$\alpha$.

\subsubsection{Far-Infrared Continuum}

A significant fraction of the bolometric luminosity of a galaxy may be
absorbed by interstellar dust and re-emitted in the thermal infrared
(10 -- 300 $\mu$m). The absorption cross-section of the dust is
strongly peaked in the UV, so to first order the far-infrared emission
scales with the star formation rate. In the limiting case of a dust
cocoon surrounding a star-forming galaxy:

\begin{eqnarray*}
  SFR (M_\odot~{\rm yr}^{-1}) = 4.5 \times 10^{-44} L_{FIR} ({\rm ergs}~{\rm s}^{-1})
\end{eqnarray*}

\subsubsection{AGN Contamination}

In many cases (particularly at higher redshifts), it is difficult to
spatially resolve the emission produced by star formation from that
produced by the AGN. In this situation, one must use spectroscopic
methods to disentangle the two processes. In some instance, one may
use the strengths of stellar atmospheric features to quantify the
starburst. These include the Balmer series, Ca I triplet
$\lambda\lambda\lambda$8498, 8542, and 8662 in the visible/deep-red
and Si IV $\lambda$1400, C IV $\lambda$1550, and He II $\lambda$1640
in the UV (Robert et al. 1993).

The emission lines produced in the ionized gas also bear the signature
of the source of energy. The ionizing spectra of all but the hottest O
stars cut off near the He II edge (54.4 eV), while AGNs are generally
strong X-ray emitters. These high-energy photons have two effects on
the emission line spectrum: (1) the material near the AGN is more
highly ionized and emit strong high-ionization lines; (2) due to the
strong energy dependence of the absorption cross-section ($\sigma_\nu
\propto \nu^{-3}$), these high-energy photons are absorbed deeper into
the gas clouds and produce extended partially ionized zones. These
zones are strong emitters of collisionally excited low-ionization
lines. One therefore expects an enhancement of {\em both} high- and
low-ionization lines in AGN relative to those in HII regions. Several
diagnostic diagrams at optical and near-infrared wavelengths have been
designed to specifically take advantage of these differences (e.g.,
Veilleux \& Osterbrock 1987; Osterbrock, Tran, \& Veilleux 1992;
Kewley et al. 2001).

In deeply obscured galaxies, the UV, optical, and near-infrared
diagnostics cannot be used to distinguish between starbursts and
AGN. One must rely on the relative intensities of the mid-infrared
fine structure lines and/or the strengths (equivalent widths) of the
polycyclic aromatic hydrocarbon (PAH) features. The principles behind
the use of the fine structure lines are roughly the same as for the
optical/UV lines {\em i.e.} use line ratio diagrams that take advantage of
the fact that AGN are copious emitters of low- and high-ionization
lines. The use of the PAH features nicely complements that of the fine
structure lines since they are generally easier to detect in the
fainter, more distant galaxies. The PAH features are less visible in
AGN due to the much stronger continuum in these objects and possible
PAH destruction.

\subsection{Evidence for a Starburst-AGN Connection}
\label{}

Since the triggering mechanism for AGN activity probably depends on
the luminosity of the AGN, I make a distinction in the following
discussion between the nearby, low-luminosity Seyferts and
Fanaroff-Riley type I (FR I) radio galaxies and the more distant and
powerful quasars, Fanaroff-Riley type II (FR II) radio galaxies, and
ultraluminous infrared galaxies (ULIRGs; log [L$_{IR}$/L$_\odot$]
$\ge$ 12 by definition - this is the subject of Lecture \#2; \S 2).

\subsubsection{Low-Luminosity AGN}

Direct evidence for recent nuclear star formation exists in a number
of Seyfert 2 galaxies ({\em i.e.} Seyferts without broad recombination
lines).  Optical and ultraviolet spectroscopy of the nuclear regions
of these galaxies often reveals the signatures of young and
intermediate-age stars.  The stellar Ca II triplet feature at
$\lambda\lambda\lambda$8498, 8542, 8662 in Seyfert 2s has an
equivalent width similar to that in normal galaxies while the stellar
Mg~Ib $\lambda$5175 is often weaker (Terlevich, Diaz, \& Terlevich
1990; Cid Fernandes et al. 2004).  This result is difficult to explain
with a combination of an old stellar population and a featureless
power-law continuum from an AGN.  The most natural explanation is that
young red supergiants contribute significantly to the continuum from
the central regions.

Evidence for intermediate-age (a few 100 Myrs) stars in Seyfert
galaxies is also apparent in the blue part of the spectrum, where the
high-order Balmer series and He I absorption lines appear to be
present in more than half of the brightest Seyfert 2 galaxies (e.g.,
Cid Fernandes \& Terlevich 1995; Joguet et al. 2001; Gonz\'alez
Delgado, Heckman, \& Leitherer 2001; Fig. 2). A few of these objects
may even harbor a broad emission feature near 4680 \AA, possibly the
signature of a population of young (a few Myrs) Wolf-Rayet stars
(Gonz\'alez Delgado et al.  2001).  The ultraviolet continuum from
some of the brightest UV Seyfert 2s also appears to be dominated by
young stars based on the strength of absorption features typically
formed in the photospheres and in the stellar winds of massive stars
(e.g., Heckman et al. 1997; Gonz\'alez Delgado et al.  1998; Fig. 2).
The bolometric luminosities of these nuclear starbursts ($\sim$
10$^{10}$ L$_\odot$) are similar to the estimated bolometric
luminosities of their obscured Seyfert 1 nuclei.  This explains why
UV-bright stellar clusters are more frequently detected in Seyfert 2s
than in Seyfert 1s (Mu\~noz Mar\'in et al.  2007). The recent
detection of near-infrared CN bands in Seyferts brings support to the
idea that star formation is indeed connected to the AGN in these
objects (Riffel et al. 2007).

\begin{figure}
\caption{($left$) UV spectrum of NGC 7130. It is displayed in log
  ($F_\lambda$) to show the emission and absorption lines. The most
  important stellar wind and photospheric absorption lines are
  labeled. ($right$) Normalized optical spectrum of NGC 7130 (dashed line)
  plotted with the normalized spectrum of a B0 V star combined with a
  G0 V star (thick line). 60\% of the light is from a B0 V and 40\%
  from a G0 V star. The comparison shows that most of the stellar
  features in NGC 7130 are well reproduced by a combination of young
  (B0 V) and old (G0 V) stars. (From Gonzalez Delgado et al. 1998)}
  \end{figure}

In recent years, SDSS has contributed significantly to our knowledge
of local AGNs. For instance, Kauffmann et al. (2003b) have studied a
sample of 22,623 narrow-line AGN with 0.02 $<$ z $<$ 0.3. They find
that the hosts of low-luminosity AGN have a stellar population
similar to that of normal early-type galaxies, while the hosts of
high-luminosity AGN have much younger mean stellar ages. Indeed,
young ($<$ 1 Gyr) stellar population appears to be a general
property of AGN with high [O~III] luminosity (Figs. 3 and 4). This
is true regardless of the presence of broad recombination lines
({\em i.e.} types 1 and 2). The young stars are spread out over
scales of at least a few kpc.

\begin{figure}
\caption{ The strengths of the 4000 \AA\ break and H$\delta$
  absorption feature are plotted as a function of log $L[O~III]$. The
  solid line shows the median, while the dashed lines indicate the
  16-84 percentiles of the 1/V$_{\rm max}$ weighted distribution.
  (From Kauffmann et al. 2003b)}
\end{figure}

\begin{figure}
\epsscale{0.5}
\caption{The fraction $F$ of AGN with H$\delta_A$ values that are
  displaced by more than 3 $\sigma$ above the local of star-forming
  galaxies is plotted as a function of log $L[O~III]$. The dashed line
  indicates the fraction of such systems in the subsample of normal
  massive galaxies. The dotted line indicates the fraction of such
  systems in the subsample of normal massive galaxies with $D_n(4000) <
  1.6$. (From Kauffmann et al. 2003b)}
\end{figure}

\subsubsection{High-Luminosity AGN}

Abundant molecular gas has been detected in radio galaxies and quasars
(e.g., Evans et al. 2001), but is this gas forming stars? The
extensive multiwavelength data set on these objects seems to indicate
that starbursts are indeed present in local and distant quasars.
Approximately 20 -- 30\% of all PG~QSOs show an infrared excess
L$_{IR}$/L$_{blue}$ $>$ 0.4.  IR-excess QSOs tend to have large dust
and H$_2$ masses, suggesting that the infrared -- submm ``bump'' in
the spectral energy distribution of PG~QSOs is due to star formation.

More direct evidence for a starburst-AGN connection has recently been
found in local quasars from Spitzer mid-infrared spectroscopy
(Schweitzer et al. 2006; Shi et al. 2007). PAH emission is detected in
11 of 26 PG QSOs and in the average spectrum of the other 15 PG QSOs
(Schweitzer et al. 2006). The strength of the PAHs in these quasars is
consistent with the far-infrared luminosity being produced primarily
by U/LIRG-like starbursts with star formation rates of order 2 -- 300
M$_\odot$ yr$^{-1}$ (Fig. 5). The strength of the starburst (measured
by the FIR or PAH luminosity) correlates with that of the QSO (based
on the 5100 \AA\ luminosity, a direct indicator of the mass accretion
rate onto the black hole; Netzer et al. 2007; Fig. 6). This suggests a
strong starburst -- AGN connection in these objects.

\begin{figure}
\epsscale{1.0}
\caption{PAH fluxes $F$(PAH 7.7 $\mu$) vs. $F$(60 $\mu$m) for local QSOs
  and starburst-dominated ULIRGs. (From Schweitzer et al. 2006)}
\end{figure}

\begin{figure}
\caption{$Top:$ Correlation of the optical (5100 \AA) and FIR (60
  $\mu$m) continuum luminosities. $Bottom:$ $L(5100)$ vs. $L$(PAH 7.7
  $\mu$m) showing detections (filled squares) and upper limits (open
  squares). (From Netzer et al. (2007)}
\end{figure}

A recent Spitzer study of high-z quasars by Maiolino et al. (2007a)
fail to detect PAH emission in these objects.  This would indicate
that the correlation between star formation rate and AGN power
``saturates'' at high luminosities (Fig. 7). The ``flattening'' of the
relation also seems to be present for CO emission (Maiolino et al.
2007b). It may therefore be that not enough fuel is available at high
$z$ to fuel the starburst and match the strength of the AGN in these
quasars.
 
\begin{figure}
\caption{ ($a$) PAH(7.7~$\mu$m) luminosity as a function of the QSO
  optical luminosity. Blue diamonds are data from Schweitzer et al.
  (2006). The red square is the upper limit obtained by the average
  spectrum of luminous, high-z QSOs.  ($b$) Distribution of the
  PAH(7.7~$\mu$m) to optical luminosity ratio in the local QSOs sample
  of Schweitzer et al. (2006). The hatched region indicates upper
  limits.  The red vertical line indicate the upper limit inferred
  from the average spectrum of luminous QSOs at high-z. (From Maiolino
  et al.  2007a) }
\end{figure}

\subsection{Discussion}
\label{}

As explained in \S 1.3, starbursts often coexist with actively accreting
supermassive black holes, but the starburst-AGN relation appears to be
tighter at high luminosity than at low luminosity. This suggests that
black hole fueling in low-luminosity AGN is more stochastic and does
not necessarily scale with the surrounding starburst.

The fueling of AGN requires mass accretion rates $\dot{M}$ $\approx$
1.7 (0.1 / $\epsilon$)(L/10$^{46}$ ergs s$^{-1}$) M$_\odot$ yr$^{-1}$,
where $\epsilon$ is the mass-to-energy conversion efficiency. A modest
accretion rate of order $\sim$ 0.01 M$_\odot$ yr$^{-1}$ is therefore
sufficient to power a Seyfert galaxy. Only a small fraction of the
total gas content of a typical host galaxy is therefore necessary for
the fueling of these low-luminosity AGNs. A broad range of mechanisms
including intrinsic processes (e.g., stellar winds and collisions,
dynamical friction of giant molecular clouds against stars; nuclear
bars or spirals produced by gravitational instabilities in the disk)
and external processes (e.g., minor galaxy interaction or mergers)
may be at work in these objects.  So it may not be surprising after
all that the power of these AGN does not necessarily scale with the
surrounding starburst.

The stringent requirements on the mass accretion rates for luminous
AGNs almost certainly require external processes such as major galaxy
interactions or mergers to be involved in triggering and sustaining
this high level of activity over $\sim$ 10$^8$ years.  Starbursts are
a natural consequence of major mergers. This topic will be discussed
in more detail in Lecture \#2 (\S 2).  A lack of ``fuel'' (= molecular
gas) may explain the apparent break in the starburst-AGN correlation
at very high AGN luminosities.

\section{Ultraluminous Infrared Galaxies}

\subsection{Introduction}
\label{}

In this lecture, I adopt the following standard definitions:
\begin{itemize}
\item Luminous infrared galaxies ($\equiv$ LIRGs) have
  log[L$_{IR}$/L$_\odot$] $\ge$ 11.0. 
\item Ultraluminous infrared galaxies ($\equiv$ ULIRGs) have
  log[L$_{IR}$/L$_\odot$] $\ge$ 12.0, 
\end{itemize}
where the infrared luminosity L$_{IR} \equiv$ L(8 -- 1000 $\mu$m). The
focus of this lecture is the ULIRGs, although we also discuss LIRGs
when relevant to our understanding of ULIRGs. The literature on ULIRGs
has been reviewed in a number of excellent articles: Sanders \&
Mirabel (1996), Blain et al. (2002), and Lonsdale, Farrah, \& Smith
(2006). These last two articles emphasize the important role that new
instrumentation [particularly SCUBA on the James Clerk Maxwell
Telescope (JCMT) and IRAC, MIPS, and IRS on the Spitzer Space
Telescope (SST)] has played in the discovery and follow-up study of
the ever-expanding population of distant ULIRGs. This short lecture
attempts to encapsulate the most important aspects of these objects.
For more information, the readers are advised to consult these three
reviews and the original papers listed in the text.

In \S 2.2, I describe the key properties of ULIRGs, both near and far.
Next (\S 2.3), I discuss the role that ULIRGs appears to play in the
formation and evolution of spheroids and their massive black holes.
ULIRGs may also contribute to the metal enrichment of the IGM through
superwinds, but the discussion of this important aspect of ULIRGs is
postponed until Lecture \#3 of this series (\S 3).

\subsection{Properties of ULIRGs: Near \& Far}
\label{}

\subsubsection{Luminosity Function}

LIRGs and ULIRGs are the dominant extragalactic population in the
local universe at $L_{BOL}$ $>$ 10$^{11}$ L$_\odot$ (Fig. 8). On
average, $\sim$ 0.008 ULIRG at $z < 0.2$ is found within each square
degree on the sky. Their number density is $\sim$ 2.5 $\times$
10$^{-7}$ Mpc$^{-3}$. The luminosity function at these high
luminosities is best described by a power law with a slope of $\sim$
--2.35.  Overall, ULIRGs are more numerous than optical selected QSOs,
although results from new optical and infrared surveys have recently
brought the number of known QSOs up closer to that of ULIRGs (e.g.,
Wisotzki 2000).

\begin{figure}
\epsscale{0.75}
\caption{ The luminosity function for infrared galaxies compared with
  other extragalactic objects. (From Sanders \& Mirabel 1996)}
\end{figure}

\subsubsection{Spectral Energy Distribution}

The spectral energy distribution of ULIRGs is dominated by a large
mid-to-far infrared bump: more than $\sim$ 80\% of the bolometric
luminosity of ULIRGs is emitted in this wavelength region (Fig. 9;
Sanders \& Mirabel 1996). This emission is due to warm (30 -- 200 K)
dust heated by a starburst and/or an AGN. A significant ($\sim$ 15\%)
subset of all local ULIRGs have ``warm'' mid-infrared colors
($f_{25}/f_{60} > 0.2$), more typical of powerful radio galaxies and
optically selected quasars. A direct comparison of the SEDs of warm
ULIRGs with those of PG~QSOs emphasizes the resemblance (Fig. 9).
The mean SED of optical selected QSOs is dominated by the ``big blue
bump'' ($\sim$ 0.05 -- 0.5 $\mu$m) and thermal emission from an
infrared/submillimeter bump ($\sim$ 1 -- 300 $\mu$m), which is
typically 30\% as strong as the big blue bump. About 20 -- 30\% of all
PG~QSOs have an infrared excess L$_{IR}$/L$_{BBB}$ $>$ 0.4. This
resemblance in SEDs has been used to argue that warm ULIRGs represent
transition objects between cool ULIRGs and QSOs (e.g., Sanders et al.
1988a,b). We return to this point in \S 2.3.

\begin{figure}
\epsscale{1.0}
\caption{($left$) Variation of the mean SEDs (from submillimeter to UV
  wavelengths) with increasing $L_{\rm IR}$ for a 60 $\mu$m sample of
  infrared galaxies. (Insert) Examples of the subset (15\%) of ULIRGs
  with ``warm'' infrared color ($f_{25}/f_{60} > 0.3$).  ($right$) Mean
  spectral energy distributions from radio to X-ray wavelengths of
  optically selected radio-loud and radio-quiet QSOs and blazars.
  (From Sanders \& Mirabel 1996)}
  \end{figure}

\subsubsection{Evolution}

Results from several studies with the Infrared Space Observatory
(ISO), Spitzer Space Telescope (SST), and SCUBA on the JCMT have
revealed a strong evolution in the luminosity function at the highest
infrared luminosities (which was hinted by the earlier {\em IRAS}
data). For instance the number density of submm galaxies (SMGs)
discovered with SCUBA at $z \sim 2$ is 10$^{2-3}$ times the number
density of local ULIRGs, {\em i. e.} similar to the number density of
bright QSOs at $z$ $\sim$ 2 (Chapman et al. 2005). This luminosity
dependence of the evolution has been confirmed with SST
(Perez-Gonzalez et al. 2005; Fig. 10).  LIRGs and ULIRGs become
important contributors to the extragalactic infrared background at $z
\sim 1$ and $z \sim 2$, respectively (Le Floc'h et al. 2005; Fig. 11).

\begin{figure}
\epsscale{0.75}
\caption{Relative contribution of starbursts ($L_{\rm IR} < 10^{11}$
  L$_\odot$), LIRGs ($L_{\rm IR} > 10^{11}$ L$_\odot$), and ULIRGs
  ($L_{\rm IR} > 10^{12}$ L$_\odot$) to the total SFR density of the
  universe as a function of redshift. Top and bottom panels show the
  two extreme cases of the faint-end slope value. The error bars show
  the uncertainties on the integration in each luminosity range,
  considering the errors of the individual luminosity function
  parameters. (From Perez-Gonzalez et al. 2005)}
\end{figure}

\begin{figure}
\epsscale{1.0}
\caption{Evolution of the comoving IR energy density up to $z \sim 1$
  (green filled region) and the respective contributions from
  low-luminosity galaxies ({\em i.e.}, $L_{\rm IR} < 10^{11}$
  L$_\odot$; blue filled area), IR-luminous sources ({\em i.e.},
  $L_{\rm IR} > 10^{11}$ L$_\odot$; orange filled region), and ULIRGs
  ({\em i.e.}, $L_{\rm IR} > 10^{12}$ L$_\odot$; red filled region).
  The solid line evolves as $(1 + z)^{3.9}$ and represents the best
  fit of the total IR luminosity density at 0 $\le z \le$ 1. Estimates
  are translated into an IR-equivalent SFR density given on the right
  vertical axis, where an absolute additional uncertainty of 0.3 dex
  should be added to reflect the dispersion in the conversion between
  luminosities and SFR. Note that the percentage of the contribution
  from each population is likely independent of this conversion. The
  dashed line corresponds to the SFR measured from the UV luminosity
  not corrected from dust extinction. The dotted line represents the
  best estimate of the total SFR density as the sum of this
  uncorrected UV contribution and the best fit of the IR-SFR (solid
  line). At $z \sim 1$ IR-luminous galaxies represent 70\% $\pm$
  15\% of the comoving IR energy density and dominate the star
  formation activity. Open diamonds and vertical and horizontal bars
  represent integrated SFR densities and their uncertainties estimated
  within various redshift bins and taken from the literature.  (From
  Le Floc'h et al.  2005)}
\end{figure}

\subsubsection{Morphology}

All but one object in a local sample of 118 ULIRGs show signs of a
strong tidal interaction/merger (Veilleux, Kim, \& Sanders 2002).
Multiple mergers involving more than two galaxies are seen in less
than 5\% of these systems.  None of the local ULIRGs is in the
first-approach stage of the interaction, and most (56\%) of them
harbor a single disturbed nucleus and are therefore in the later
stages of a merger (Fig. 12).  The fraction of post/old mergers ({\em
  i. e.}  single-nucleus systems) increases with infrared luminosity.
This trend with luminosity is seen within the class of ULIRGs but is
even more obvious when combining LIRGs with ULIRGs (Ishida 2004;
Veilleux et al.  2002): this fraction is $\sim$ 10\% (20\%) among
LIRGs with log[L$_{IR}$/L$_\odot$] = 11.25 -- 11.50 (11.5 - 11.75),
increases to $\sim$ 50\% among ULIRGs with log[L$_{IR}$/L$_\odot$] =
12.0 -- 12.2, and peaks at $\sim$ 80\% among the most luminous
(log[L$_{IR}$/L$_\odot$] $>$ 12.5) ULIRGs.

\begin{figure}
\epsscale{0.75}
\caption{Apparent nuclear separations in the 1-Jy sample of galaxies.
  The distribution is highly peaked at small values but also presents
  a significant tail at high values. The very uncertain separation
  measured in F11223$-$1244 (87.9 kpc) is not shown in this figure.
  (From Veilleux et al. 2002)}
\end{figure}

Distant ULIRGs also often appear to be undergoing a merger. Most
high-z SMGs (50 -- 60\% at optical wavelengths and 85\% in the UV) are
multi-component or disturbed systems, suggestive of mergers or
interactions (e.g., Chapman et al. 2003; Smail et al. 2004).

\subsubsection{Gas Content}

ULIRGs are extremely rich in molecular gas with H$_2$ masses of a few
$\times$ 10$^{10}$ M$_\odot$, more than an order of magnitude that of
the Milky Way galaxy (e.g., Sanders et al. 1988c). The physical
conditions in this gas are similar to those in massive giant molecular
cloud (GMC) cores in our own Galaxy, although they are slightly more
infrared luminous per unit of molecular gas mass and denser than GMC
cores (Solomon et al. 1997).  Remarkably about 40 -- 10\% of this
molecular gas lies within the central kpc of ULIRGs. This implies
molecular gas surface densities of $\sim$ few $\times$ 10$^4$
M$_\odot$ pc$^{-2}$, similar to the stellar densities of the core of
elliptical galaxies (e.g., Downes \& Solomon 1998). This large
concentration of activity in ULIRGs is also seen at mid-infrared
wavelengths (Soifer et al. 2000). These results are a natural
consequence of merger-induced activity: in modern simulations of
galaxy mergers the gas originally distributed over the entire body of
each galaxy is funnelled rapidly to the central kpc of the merger due
to bar-induced torques and energy dissipation, triggering a powerful
nuclear starburst and possibly fueling an AGN (e.g., Barnes \&
Hernquist 1996; Mihos \& Hernquist 1996; Hopkins et al 2006).

\subsection{Importance of ULIRGs}
\label{}

\subsubsection{Spheroids in Formation}

The end-result of the simulations discussed in \S 2.2.5 is an
elliptical-like merger remnant, but do observations support this
picture? To answer this question careful imaging and spectroscopic
studies of ULIRG are needed to derive the morphology and kinematics of
the underlying host galaxies, and compare the results with those of
normal ellipticals.  In Veilleux et al. (2006), we found that the
removal of the central PSF emission associated with the AGN or nuclear
starburst is an important source of errors in the analysis of the
surface brightness profiles in the more nucleated ULIRGs ({\em i. e.}
those in the later stage of a merger).  A detailed two-dimensional
analysis of the surface brightness distributions in these objects
indicates that the great majority (81\%) of the single-nucleus systems
show a prominent early-type morphology.  As shown in Fig. 13, the hosts
of ULIRGs lie close to the locations of intermediate-size ($\sim$ 1 --
2 $L^*$) spheroids in the photometric projection of the fundamental
plane of ellipticals, although there is a tendency for the ULIRGs with
small hosts to be brighter than normal spheroids.  Excess emission
from a merger-triggered burst of star formation in the ULIRG hosts may
be at the origin of this difference.

\begin{figure}
\caption{Surface brightnesses vs. half-light radii for the early-type
  host galaxies in the sample of Veilleux et al. (2006). The hosts of
  the seven PG QSOs in the sample are statistically indistinguishable
  from the hosts of the 1 Jy ULIRGs. Both classes of objects fall near
  the photometric fundamental plane relation of elliptical galaxies as
  traced by the data of Pahre (1999; dashed line), although the
  smaller objects in our sample tend to lie above this relation (the
  solid line is a linear fit through our data points). This may be due
  to excess H-band emission from a young stellar population. ULIRGs
  and PG QSOs populate the region of the photometric fundamental plane
  of intermediate-size ($\sim$ 1 -- 2 L$^*$) elliptical/lenticular
  galaxies.  In contrast, the hosts of the radio-bright quasars of Dunlop
  et al.  (2003) are massive elliptical galaxies that are
  significantly larger than the hosts of ULIRGs and PG QSOs. For this
  comparison, the R-band half-light radii tabulated in Dunlop et al.
  were taken at face value, and the surface brightnesses in that paper
  were shifted assuming $R - H = 2.9$, which is typical for early-type
  systems at $z \sim 0.2$.  (See Veilleux et al. 2006 for more
  detail)}
\end{figure}

VLT/Keck near-infrared stellar absorption spectroscopy has also been
carried out to constrain the host dynamical mass for many of these
ULIRGs. The analysis of these data (Dasyra et al.  2006ab) builds on
the analyses of Genzel et al. (2001) and Tacconi et al.  (2002) and
reveals that the majority of ULIRGs are triggered by almost equal-mass
major mergers of 1.5:1 average ratio, in general agreement with
Veilleux et al. (2002). In Dasyra et al., we also find (see Fig. 14)
that coalesced ULIRGs resemble intermediate mass
ellipticals/lenticulars with moderate rotation, in their velocity
dispersion distribution, their location in the fundamental plane (FP;
e.g., Kormendy \& Djorgovski 1989) and their distribution of the ratio
of rotation/velocity dispersion [v$_{\rm rot}$ sin(i)/$\sigma$]. These
results therefore suggest that ULIRGs form moderate mass ($m^\ast \sim
10^{11}$ M$_\odot$), but not giant (5 -- 10 $\times$ 10$^{11}$
M$_\odot$) ellipticals. These results are largely consistent with
those from our imaging studies.

\begin{figure}
\caption{$R_{eff}$ - $\sigma$ projection of the early-type galaxy FP.
  For viewing clarity, the various types of mergers are plotted in
  separate panels. The ULIRG remnants are plotted as triangles (left
  panel).  The LIRGs and other (visually selected) mergers are
  plotted as diamonds and open-crossed diamonds, respectively (right
  panel). The effective radii of all merger remnants used in this
  figure are equal to the averages of their NIR measurements, if more
  than one is available. (See Dasyra et al. 2006b for more detail)}
\end{figure}

A similar analysis of the distant ULIRGs is obviously much more
difficult. Current results suggest that high-z submm galaxies are on
average bigger and more gas-rich systems than local ULIRGs (see Table
1 for a head-to-head comparison). The activity in these objects
appears to be more extended than in local ULIRGs, so the high-z
ULIRGs may not be simply scaled-up versions of local ULIRGs (Chapman
et al. 2003).

\vskip 0.1in

\begin{table}[htbp]
\centerline{Table 1. Comparison of Distant Submm Galaxies with Local ULIRGs}
\vskip 0.1in
\begin{center}
\begin{tabular}{llll}
\hline
\hline
Property & Submm Galaxies & Local ULIRGs\\
\hline
$<v_c>$ & 400 km s$^{-1}$ & 240 km s$^{-1}$ \\
$<M_{dyn,1/2}>$ & 7 $\times$ 10$^{10}$ M$_\odot$ & 5 $\times$ 10$^9$ M$_\odot$\\
$<R_{1/2}>$ & 2.0 kpc & 0.6 kpc \\
$<L_{BOL}>$ & 10$^{13.1}$ L$_\odot$ & 10$^{12.2}$ L$_\odot$ \\
$<M_{gas}/M_{dyn}>$ & 0.3 -- 0.4 & 0.16 \\
$<\Sigma_{dyn}>$ & 5000 M$_\odot$ pc$^{-2}$ & 4900 M$_\odot$ pc$^{-2}$ \\
\hline
\end{tabular}
\end{center}
\end{table}

A simple exercise in ``numerology'' quickly shows that local ULIRGs
cannot be an important contributors of early-type galaxies: the number
density of local ULIRGs $\sim$ 2.5 $\times$ 10$^{-7}$ Mpc$^{-3}$
$\sim$ (1/7000) $\times$ number density of SDSS ellipticals. The story
at high redshifts is different. The current best estimates for the
stellar masses and dynamical masses in the cores of high-$z$ ULIRGs
are $\sim$ 10$^{11}$ M$_\odot$. The ULIRG lifetime of these objects,
estimated from stellar population analyses, is $\sim$ 200 -- 300 Myr,
{\em i.e.} of the same order of magnitude as the gas consumption time
scale ($\sim$ 40 Myr). The ULIRG era spans $z \sim 1.5 - 3$,
corresponding to a time scale of $\sim$ 1.5 Gyr (this number is quite
uncertain; here we have choosen a conservatively low number). So the
expected number density of descendants from these high-$z$ ULIRGs is
$n$(descendants) = $n$(bright SMGs) $\times$ [$\tau$(ULIRG era) /
$\tau$(ULIRG lifetime)] $\sim$ 3 $\times 10^{-5}$ Mpc$^{-3}$ (1500/300)
$\sim 1.5 \times 10^{-4}$ Mpc$^{-3}$ $\sim n$($> L^*$ ellipticals at
$z \sim 0$). The measured average absolute magnitude of these high-z
ULIRGs ($<M_K> \sim -26.4$ at $z \sim 2.2$) is consistent with that of
$>$ $L^*$ elliptical at $z \sim$ 0, after taking into account the
fading of the stellar population.

\subsubsection{Black Hole Growth}

The fraction of AGN (Seyfert nuclei) has long been known to increase
with infrared luminosity, from about a third among ULIRGs with
log[L$_{IR}$/L$_\odot$] $\sim$ 12 to about a half among ULIRGs with
log[L$_{IR}$/L$_\odot$] $\ge$ 12.3 (e.g., Veilleux et al. 1995,
1999ab; Fig. 15). This trend is also seen at mid-infrared wavelengths
(Lutz et al. 1999; Tran et al. 2001; Veilleux et al. 2008).

\begin{figure}
\caption{The optical spectral classification of infrared galaxies
  versus infrared luminosity (From Veilleux et al. 1995, 1999a)}
\end{figure}

A detailed morphological study on local ULIRGs and PG~QSOs conducted
by our group (Veilleux et al. 2006) shows that (1) nearly all
quasar-like ULIRGs are advanced mergers, (2) starburst-like ULIRGs are
found in all merger phases, (3) warm, AGN-like ULIRGs live in
early-type hosts, (4) tidal features are weaker among warm, AGN-like,
early-type ULIRGs, and (5) the host sizes and luminosities of PG~QSOs
are statistically indistinguishable from those of the ULIRG hosts [in
comparison, radio brighter quasars, such as those studied by Dunlop et
al. (2003), have hosts which are larger and more luminous; Fig. 13].
All of these results, except \#2, provide support for a possible
merger-driven evolutionary connection between cool ULIRGs, warm
ULIRGs, and PG~QSOs. However, this sequence may break down at low
luminosity since the lowest luminosity PG~QSOs in the sample of
Veilleux et al. (2006) show distinct disk components which preclude
major (1:1 -- 2:1) mergers.  The presence of starburst-like ULIRGs in
all merger phases seems to indicate that this merger scenario is not
100\% efficient at producing an AGN.

A kinematic analysis of a dozen PG~QSOs by our group (Dasyra et al.
2007) shows agreement between the host mass (thus black hole mass) of
PG~QSOs and coalesced ULIRGs (see Table 2).  Converting the host
dispersion in fully coalesced ULIRGs into black hole mass with the aid
of the M$_{\rm BH} - \sigma$ relation (e.g., Gebhardt et al. 2000)
yields black hole mass estimates of the order $10^7 - 10^8$ M$_\odot$.
The accretion rate for sources after the nuclear coalescence is high
$0.5 - 0.9$, similar to those derived by Veilleux et al. (2002,
2006). 

\vskip 0.1in

\begin{table}[htbp]
\centerline{Table 2. Comparison of Local QSOs with Local ULIRGs}
\vskip 0.1in
\begin{center}
\begin{tabular}{ccccc}
\hline
\hline
Class & log[L$_{BOL}$/L$_\odot$] & $m/m^*$ & log $m_{BH}$ & $\eta_{Edd}$\\
      &  ... & ($m^*$ = 1.4 $\times$ 10$^{11}$ M$_\odot$) &
      (M$_\odot$) & \\
\hline
$<$ULIRG$>$ & 12.2 & 0.8 & 7.9 & 0.5 \\
$<$PG QSOs$>$ & 12.2 & 1.5 & 8.0 & 0.3\\
\hline
\end{tabular}
\end{center}
\end{table}

The situation at high redshifts is much less clear. Most ($\sim$ 50 --
75\%) submm galaxies with log[L$_{IR}$/L$_\odot$] $>$ 12.5 harbor an
AGN (Seyfert nucleus; Smail et al. 2002; Chapman et al. 2005), but
apparently the AGN generally does not dominate the energy output in
these systems (e.g.  Alexander et al. 2005). This AGN fraction appears
to increase with infrared luminosity, as in the case of their lower
redshift counterparts (Sajina et al. 2007). The black hole mass in
these systems is essentially unknown, except for X-ray estimates.  At
face value, the black hole to stellar mass ratios of high-z submm
galaxies appears to be 1-2 orders of magnitude smaller than the local
value, but this is only true if the black holes are accreting at the
Eddington limit and are not Compton thick. This discrepancy is reduced
by a factor of $\sim$ 5 if the virial black-hole mass estimator is
used, implying accretion at $\sim$ 20\% of the Eddington limit
(Alexander et al. 2008).

\subsection{Summary}
\label{}

The large body of data on local (distant) ULIRGs suggests that many of
them are undergoing a major merger and are in the process of forming
spheroids of intermediate (large) masses at $z \sim$ 0 (2).  The $>$
L$^*$ ellipticals we see today may be the descendants of $z \sim 2$
ULIRGs.  Black hole growth is clearly taking place along the merger
sequence, and the results of detailed studies at low redshift provide
support for an evolutionary connection between ULIRGs and {\em some}
QSOs.  However, it is not clear at present whether the end-result of
high-z ULIRGs are indeed quasars. This evolutionary scenario requires
that an important fraction of the gas mass be removed from the cores
of these objects to reveal the QSOs. The presence of galactic winds
may therefore be an important condition for this evolutionary scenario
-- this is the subject of Lecture \#3 (\S 3).

\section{Feedback Processes and their Impact on Galaxy Formation and Evolution}

\subsection{Introduction}
\label{}

Feedback processes likely have a strong impact on the formation of
galaxies and their evolution of galaxies. These processes are divided
into two broad categories: mechanical and radiative. In the first
case, mechanical energy or momentum from starbursts and/or AGN affects
the thermal and chemical properties of galaxies over the entire galaxy
mass spectrum. Radiative feedback may take several forms.  Radiation
pressure may be important near luminous starbursts and/or AGN.
Radiative heating may also be dynamically important in luminous AGN,
causing Compton-heated winds in the cores of these objects. Radiation
may also ionize the surrounding ISM.  Ionization cones in local AGN
(Wilson \& Tsvetanov 1994) and the proximity effect near QSOs (e.g.,
Bajtlik, Duncan, \& Ostriker 1988) are direct consequences of this
process. Radiation may also destroy molecules (H$_2$, polycyclic
aromatic hydrocarbons [PAHs]) and cause dust grains to sublimate.
This process is particularly important in dwarf galaxies at high
redshifts.

The main focus of this lecture is mechanical feedback.  Galactic winds
are the primary mechanism by which energy and metals are recycled in
galaxies and are deposited into the intergalactic medium.  New
observations are revealing the ubiquity of this process, particularly
at high redshift.  Here, I first describe the basic physics behind
winds (\S 3.2), discuss the observational evidence for them in nearby
star-forming and active galaxies (\S\S 3.3 and 3.4) and in the
high-redshift universe (\S 3.5), and consider the implications of
energetic winds for the formation and evolution of galaxies (\S 3.6).
Finally, in \S 3.7, I describe a series of unanswered questions which
could possibly be tackled as PhD thesis projects. Much of the
discussion is inspired from the review article by Veilleux, Cecil, \&
Bland-Hawthorn (2005) with recent updates when needed.  For more
detail on high-z winds, the readers should refer to Ferrara (2007).

\subsection{Basic Physics of Mechanical Feedback}
\label{}

There are two possible sources of energy for mechanical feedback in
galaxies: starbursts or AGNs. In starburst winds, the mechanical
energy / momentum is provided either by stellar (OB, WR) winds or by
supernovae. Their contributions are expected to scale with the star
formation rate. In an instantaneous starburst, stellar winds are
expected to dominate during the first $\sim$ 6 $\times$ 10$^6$ yrs and
then SNe take over until $\sim$ 40 Myr. AGN winds, on the other hand,
may be driven by radiation pressure, radiative heating (Compton-heated
wind; Begelman 1985), or magnetic fields anchored in the accretion
disks. In this last case, the outflow is in the form of loosely (or
highly) collimated jets depending on whether one is dealing with a
radio-quiet (or radio-loud) AGN. The contributions to the winds are
expected to scale roughly with the luminosity of the AGN or the mass
accretion rate onto the supermassive black hole.

Thermalization of the mechanical energy by the starburst or the AGN
creates a cavity of hot gas in the starburst or near the AGN with a
temperature
\begin{equation}
T = 0.4~\mu~m_H~\dot{E}/k\dot{M},
\end{equation}
where $\dot{E}$ is the fraction of the mechanical energy injection
rate that is thermalized and $\dot{M}$ is the rate at which the mass
is heated. For a starburst, equation (1) becomes:
\begin{equation}
T \approx 3 \times 10^8~\xi~\Lambda^{-1}~~{\rm K},
\end{equation}
where $\xi$ is the thermalization efficiency of the mechanical energy.
The parameter $\Lambda$ is the mass-loading factor, the ratio of the
total mass of heated gas to the mass that is directly ejected by SNe
and stellar winds or by the AGN. It accounts for the possibility that
some of the ISM is mixed with the stellar or AGN ejecta. Note that
this tenuous hot gas will be a poor X-ray (bremsstrahlung) emitter
unless $\xi/\Lambda \ll 1$.

The pressure created by this hot gas can significantly exceed the
pressure of the undisturbed ISM, hence drive the bubble outflow.
Bubble evolution in gas-rich systems is described by the self-similar
Taylor-Sedov solutions to a point-source explosion (blast wave) in a
homogeneous medium (Taylor 1950; Sedov 1959).  If radiative losses of
the overall system are negligible, the expanding bubble is {\em
  energy-conserving} and the velocity of the expanding shell of
shocked ISM are given by (Castor, McCray, \& Weaver 1975; Weaver et
al. 1977)
\begin{equation}
V_{\rm shell} = 640~(\xi\dot{E}_{44}/n_o)^{1/5}~t_6^{-2/5} = 670~(\xi\dot{E}_{44}/n_o)^{1/3}~r_{\rm shell, kpc}^{-2/3}~~{\rm km~s^{-1}},
\end{equation}
where $t_6$ is the age of the bubble in Myr, $n_o$ is the ambient
density in cm$^{-3}$, and $\dot{E}_{44}$ is the mechanical luminosity
of the wind in units of 10$^{44}$ erg~s$^{-1}$. Once the shell has
formed, if the wind-blown bubble approaches the scale height of the
disk $H$, the shell reaccelerates, begins to fragment through growing
Rayleigh-Taylor instabilities, and finally vents these fragments
and the freely flowing and shocked wind into the galaxy halo (e.g.,
Fig. 16).  The terminal velocity of the wind can be estimated by
equating the total energy deposition rate $\xi\dot{E}$ to the
asymptotic rate of kinetic energy loss: $\frac{1}{2}~\Lambda
\dot{M} V_{\infty}^2 \approx \xi \dot{E}$.  For a starburst-driven
wind, we obtain
\begin{equation}
V_{\infty} \approx (2~\xi \dot{E}/\Lambda \dot{M})^{\frac{1}{2}} \approx 3000~(\xi/\Lambda)^{\frac{1}{2}}~~{\rm km~s}^{-1}.
\end{equation}
The $\Lambda$ dependence is easy to understand: cold ISM gas that
feels the full brunt of the wind is shock heated and evaporated and
eventually mass loads the hot flow, which slows the wind.
Equation (4) assumes negligible halo drag.

\begin{figure}
\caption{An example of a state-of-the-art 3D hydrodynamical simulation
  of a starburst outflow in an interstellar media with fractal size
  distribution at three epochs after the constant energy input wind
  starts to blow: 1.0 Myr (bottom), 2.5 Myr (middle), and 4.0 Myr
  (top). Blue is log-density and red is log-temperature. (From Cooper
  et al. 2007)}
\end{figure}

More relevant to the measured velocity of the line-emitting debris in
the winds is the expected terminal velocity of clouds accelerated by
the wind ram pressure
\begin{equation}
V({\rm cloud}) \approx 600~\dot{p_{34}}^{0.5}~\Omega_W^{-0.5}~r_{{\rm 0,kpc}}^{-0.5} N_{{\rm cloud, 21}}^{-0.5}~~{\rm km~s}^{-1}, 
\end{equation}
where $\dot{p}_{34}$ is the wind momentum flux in units of $10^{34}$
dynes, $N_{{\rm cloud, 21}}$ is the column density of the cloud in
units of $10^{21}$ cm$^{-2}$, $r_0$ is the initial radius in kpc where
the cloud is injected into the wind, $\Omega_W$ is the solid angle of
the wind in steradians.

\subsection{A Few  Local Examples}
\label{}

\subsubsection{The Milky Way Galaxy}

\begin{figure}
\epsscale{0.5}
\caption{Aspects of the Milky Way's wind. (Top) ROSAT 1.5 keV diffuse
  X-ray map that shows a biconical pattern emerging from the Galactic
  Center on scales of tens of degrees. (Bottom) The inner
  2.5×2.5$^\prime$ around the Galactic Center. Above, the plane in red is
  the Galactic Center Lobe, here imaged by Law \& Yusef-Zadeh with
  rasters from the Green Bank Telescope. Elsewhere, the color image
  comes from 8.3 (B), 13 (G), and 21.3 (R) $\mu$m scans obtained with
  the SPIRIT III radiometer on the MSX spacecraft. These show embedded
  dust at various temperatures. Note the warm dust filaments along the
  edges of the GCL. (From Veilleux et al. 2005)}
\end{figure}

Evidence is mounting for the existence of a nuclear wind in our own
Milky Way Galaxy. Sofue \& Handa (1984) discovered the 200-pc diameter
Galactic Center radio lobe (GCL, Fig.\ 17), with an implied thermal
energy of $\sim3\times 10^{51}$ erg.  Bland-Hawthorn \& Cohen (2003)
detected the GCL at mid-IR wavelengths (Fig.\ 17). The association
of the lobe with denser material raises the energetics to
$10^{54}/\kappa$ erg, where $\kappa$ is the covering fraction of the
dense shell; less energy is needed if PAHs contribute significantly to
the mid-iIR continuum emission.  These energetics assume a shell
velocity of $\sim$ 150 km~s$^{-1}$, a value based on the kinematics of
the neighboring molecular gas after correction for bar streaming; this
value is uncertain because of our location in the plane.  The ROSAT
1.5 keV diffuse X-ray map over the inner 45$^\circ$ provides
compelling evidence for this galactic wind interpretation (Fig.\ 17)
(Bland-Hawthorn \& Cohen 2003).

\subsubsection{M82}

\begin{figure}
\epsscale{1.0}
\caption{M82, imaged by the Wisconsin Indiana Yale NOAO telescope in
  H$\alpha$ (magenta) and HST in BVI continuum colors. Several of the
  largest scale filaments trace all the way back to super-starclusters
  embedded in the disk. (Courtesy Smith, Gallagher, \& Westmoquette)}
\end{figure}

M82 hosts arguably the best studied example of a galactic wind (Fig.
18).  The bright H$\alpha$-emitting filaments above and below the disk
of this object are moving at a deprojected velocity of 525 -- 655 km
s$^{-1}$.  These filaments have a total mass of $\sim$ 5.8 $\times$
10$^6$ M$_\odot$ and kinetic energy of $\sim$ 2.1 $\times$ 10$^{55}$
ergs, or about 1\% of the total mechanical energy input from the
starburst (Shopbell \& Bland-Hawthorn 1998). Deep H$\alpha$ and X-ray
images of M82 reveal that the outflow extends at least 12 kpc from the
nucleus, some of which may be escaping from the galaxy (e.g., Devine
\& Bally 1999; Lehnert et al. 1999; Stevens, Read, \& Bravo-Guerrero
2003).  Walter, Weiss, \& Soville (2002) also found that $\sim$ 3
$\times$ 10$^8$ M$_\odot$ of molecular gas is taking part in this
outflow, corresponding to a kinetic energy $\sim$ 3 $\times$ 10$^{55}$
ergs, very similar to that of the bright H$\alpha$ filaments.

\subsubsection{NGC 3079}

\begin{figure}
\caption{NGC 3079 imaged with HST (red for H$\alpha$ + [N II], green
  for I-band) and X-ray (blue). (a) Large-scale emission across 15
  $\times$ 5 kpc.  Numerous H$\alpha$ filaments rise above the disk.
  Note the V-shaped wind pattern extending in X rays from the nucleus;
  for clarity, we have suppressed the diffuse X-ray emission across
  the superbubble, where it is generally clumped (c). (b) The 1
  $\times$ 1.2 kpc superbubble in H$\alpha$ + [N II] emission, with
  log-scaled intensities. It is composed of four vertical towers of
  twisted filaments; the towers have strikingly similar morphologies.
  (c) Close-up of the wind-swept, circumnuclear region. Note how X-ray
  emission (blue) clumps along the optical filaments of the
  superbubble at the limit of {\em Chandra}'s resolution. A prominent
  dust filament at left drops out of the wind. (From Veilleux et al.
  2005) }
\end{figure}

The best known example of a ``young'' outflow is found in NGC 3079,
where a partially ruptured 1-kpc superbubble is powered by a nuclear
starburst (Fig. 19). The outflow velocity of the warm ionized gas is a
record-setting $\sim$ 1500 km s$^{-1}$.  The superbubble is composed
of 4 vertical towers of twisted filaments; the towers have strikingly
similar morphologies and are made of gas in vortex motion like that of
a mushroom cloud from an atomic bomb explosion (Veilleux et al. 1994;
Cecil et al. 2001; Cecil, Bland-Hawthorn, \& Veilleux 2002).

\subsubsection{Hot Wind Fluid}

Chemically-enriched wind fluid has been directly observed in X-rays in
very few objects: the Milky Way (Yuasa et al. 2007), M82 (Griffiths et
al. 2000; Stevens et al. 2003), and possibly NGC~1569 (Martin,
Kobulnicky, \& Heckman 2002). The fairly modest alpha-element to iron
ratio seems to imply substantial mixing of the supernova-processed
material with the ambient ISM.

\subsubsection{Dust Outflow}

\begin{figure}
\caption{Color-composite image of M82, displayed with a logarithmic
  stretch. The blue channel is the IRAC 3.6 $\mu$m image and the green
  channel is the IRAC 8.0 $\mu$m image (= PAH/dust emission), while the
  red channel is the MIPS 24 $\mu$m image, where the bright nucleus
  and associated diffraction spikes have been subtracted; the radial
  red streaks are residuals of the subtraction. The panel is
  9.2$^\prime$ $\times$ 9.2$^\prime$ on a size, or 9.6 $\times$ 10.7
  kpc. (From Engelbracht et al. 2006)}
\end{figure}

There is now direct evidence that dust also participates in
galactic-scale outflows. The best example so far is that of M82 seen
by GALEX (Hoopes et al. 2005) and Spitzer (Engelbracht et al. 2006;
Fig. 20). Galactic winds therefore provide another way (other than
galaxy interactions, for instance) to bring dust out of galaxies and
into the intergalactic medium.

\subsection{Statistics on Winds at z $<$ 0.5}
\label{}

Here we summarize the properties of local galactic winds. See Veilleux
et al. (2005) for more detail.

\begin{itemize}
\item Starburst-driven winds require a star formation rate above
  $\sim$ few M$_\odot$ yr$^{-1}$ or a star formation rate surface
  density (averaged over the optical size of the galaxy) larger than
  $\sim$ 0.001 M$_\odot$ yr$^{-1}$ kpc$^{-2}$.
\item AGN-driven winds are seen in most radio-loud AGN and $\ge$ 1/3
  of all radio-quiet AGN.
\item They subtend a solid angle $\Omega_{\rm wind} / 4 \pi \sim 0.1 -
  0.5$
\item They extend to a radius of a few to more than 50 kpc.
\item The deprojected outflow velocities range from $\sim$ 25 km
  s$^{-2}$ in dwarf galaxies to more than 1000 km s$^{-1}$ in the more
  powerful starbursts. This velocity depends on the star formation
  rate (Fig. 21) and the gas phase under consideration (it increases
  with the temperature of the gas).
\item The mass outflow rates in starbursts dM/dt$_{\rm outflow} =
  \beta$ SFR where the entrainment efficiency $\beta \sim 0.01 - 5$
  and ``saturates'' at the highest SFR.
\item The mass outflow rates in AGNs exceeds the mass accretion
  rates onto the SMBH. 
\item The escape fraction of the gas is poorly constrained. It depends
  on the temperature of the gas and the importance of the halo
  drag. Neglecting halo drag, current numbers for the outflowing
  neutral gas are $\sim$ 5 -- 20\%. It is likely higher for the hotter
  gas phase.
\item The kinetic energy involved in starburst-driven outflows is 10$^{53}$ --
  10$^{59}$ ergs $\sim$ (10 -- 50\%) of the total kinetic energy
  returned to th ISM by the starburst. It scales roughly with the star
  formation rate. 
\item The kinetic energy involved in AGN-driven outflows is similar to
  that starburst-driven outflows. It scales with the power of the AGN.
  A rough rule of thumb for the jet mechanical luminosity is
  $\sim$ 0.1 $L_{\rm Eddington}$.
\item The energy stored in radio lobes of powerful radio galaxies
  reaches $\sim$ 10$^{58}$ -- 10$^{61}$ ergs.
\end{itemize}

\begin{figure}
\caption{Maximum Na I D absorption-line outflow velocities as a
  function of (a) circular velocities and (b) star formation rates.
  Orange skeletal stars represent star-forming dwarfs from Schwartz \&
  Martin (2004), and red open stars represent infrared-selected
  starbursts from Rupke, Veilleux \& Sanders (2005a,b). Filled blue
  circles and filled green square represent Seyfert 2s and Seyfert 1s,
  respectively, from Rupke, Veilleux \& Sanders (2005c). The dashed
  line in panel a represents the escape velocity for a singular
  isothermal sphere with $r_{\rm max}/r$ = 10, whereas the dashed
  lines in panel b are characteristic velocities of ram-pressure
  accelerated clouds (Murray et al. 2005) for column densities of
  10$^{20}$ cm$^{-2}$ (top line) and 10$^{21}$ cm$^{-2}$.  (From
  Veilleux et al. 2005)}
\end{figure}

\subsection{High-z Winds}
\label{}

There is evidence for winds in the spectra of several $z >$ 1
galaxies.  Low-ionization interstellar absorption lines that are
blueshifted by hundreds of km~s$^{-1}$ relative to systemic
velocities, and Ly$\alpha$ emission lines similarly shifted redward,
have been detected in most $z \sim 3 - 4$ Lyman break galaxies (LBGs),
in several gravitationally lensed Ly$\alpha$-emitting galaxies at $z
\sim 4 - 5$, and in many luminous infrared galaxies at $z \ge
2$. Ly$\alpha$ emission with red asymmetric or P Cygni-type profiles
is also commonly seen in $z \ge 5$ Ly$\alpha$-emitting galaxies.
Outflow velocities of $\sim$ 300 km s$^{-1}$ are typically observed in
LBGs (e.g., Shapley et al. 2003).  This value is slightly higher than
the outflow velocities found in low-$z$ galaxies of similar star
formation rates.  Spectral analyses of background QSOs with
line-of-sights near the LBGs provide new constraints on the
environmental impact of LBG winds. Adelberger et al. (2005) found
significant HI deficits within 1 Mpc of $\sim$ 1/3 of LBGs.  These
results are qualitatively consistent with the idea that winds emerge
along paths of least resistance out of 500 kpc, avoiding large-scale
filaments.

\subsection{Impact on Galaxy Formation and Evolution}
\label{}

\begin{figure}
\caption{ The mass function of dark matter haloes predicted for
  various CDM cosmologies (light broken lines). The bold lines show
  the mass function of galactic haloes. Note the large discrepancy in
  slopes at small and large masses. (from Somerville \& Primack 1999)}
\end{figure}

There is growing evidence that galactic winds have inhibited early
star formation and have ejected a significant fraction of the baryons
once found in galaxies.  The latter may explain why few baryons are in
stars ($\Omega_*/\Omega_b \sim 0.1$; Fukugita, Hogan, \& Peebles 1998)
and why galaxies like the Milky Way contain fewer baryons than
expected from hydrodynamical simulations (Silk 2003). We review in
this section the impact of winds on galaxies and on their environment.

\subsubsection{Galaxy Luminosity Function}

Galactic winds have modified substantially the shape of the galaxy
luminosity function, flattening its faint-end slope compared to that
of the halo mass function (Fig. 22).  Significant feedback also appears
necessary to avoid the `cooling catastrophe' at high redshift that
would otherwise overproduce massive luminous galaxies (Fig. 22).
Energies of a few $\times~10^{49}$ ergs per solar mass of stars formed
can explain the sharp cutoff at the bright end of the luminosity
function.  Starburst-driven winds are too feeble by a factor of
several to fully account for the cutoff. It is therefore argued that
feedback from black hole accretion is the only way to expel winds hot
enough to prevent subsequent gas recapture by group halos. Feedback
from starburst- and AGN-driven winds may help set up the bi-modality
observed in galaxy properties (blue \& red sequences). AGN feedback
may be particularly effective in clustered environment where the
infalling gas is heated by a virial shock and thus more dilute.

\subsubsection{Black Hole -- Spheroid Connection}

As discussed in Lecture \# 1 (\S 1), a mechanism is needed to regulate
the growth of the black hole and the starburst and produce the black
hole -- spheroid mass relation.  Negative feedback from AGN- and/or
starburst-driven winds may be responsible for stopping the gas flow to
the AGN and the surrounding starburst (e.g., Murray et al. 2005).

\subsubsection{Mass -- Metallicity Relation}

Galactic winds may help explain the well-known galaxy mass --
metallicity relation. In this scenario, massive galaxies with deep
gravitational potentials are expected to retain more of their
supernova ejecta than dwarf galaxies.  An analysis of the Sloan
Digital Sky Survey (SDSS) database by Tremonti et al. (2004) has shown
that the gas-phase metallicity of local star-forming galaxies
increases steeply with stellar mass from 10$^{8.5}$ to 10$^{10.5}$
M$_\odot h^{-2}_{70}$, but flattens above 10$^{10.5}$ M$_\odot
h^{-2}_{70}$.  Similar trends are seen when internal velocity or
surface brightness is considered instead of stellar mass (Kauffmann et
al. 2003a). The stellar mass scale of this flattening coincides roughly
with the dynamical mass scale of metal retention derived by Garnett
(2002).  These results suggest that the chemical evolution of galaxies
with $v_c \ge 125$ km s$^{-1}$ is (virtually) unaffected by galactic
winds, whereas galaxies below this threshold tend to lose a large
fraction of their supernova ejecta.

\subsubsection{Influence on Intergalactic Scales}

Galaxy winds from early galaxies have been proposed to explain the
``entropy floor'' of galaxy clusters, the metal abundances of the
intracluster and intergalactic media, and the lack of massive cooling
flows in rich galaxy clusters with (once) powerful radio galaxies. 

\subsection{Open Issues}
\label{}

The questions in this section are directly taken from Veilleux et al.
(2005) but they are updated to take into account papers published in
the past three years. Each of these questions could be a worthwhile
PhD thesis project.

\subsubsection{Theoretical Challenges}

\begin{itemize}
\item[1.] Modeling the energy source. Current simulations do a poor
  job of modeling the energy source itself, especially in AGN-driven
  winds where energy and momentum injection rates and their dependence
  on mass accretion rate are virtually unknown.
\item[2.] Modeling the host ISM. The recent work of Cooper et al.
  (2007) is the first of a new generation of simulations able to
  handle a multiphase ISM with a broad range of densities and
  temperatures.  Such sophistication is crucial to understanding and
  predicting the mass of gas entrained in winds.
\item[3.] Coupling the radiation field to the gas. Current simulations
  do not account for possible coupling between the wind material and
  the radiation field emitted by the energy source or the wind itself,
  and indeed ignore radiation pressure.
\end{itemize}

\subsubsection{Observational Challenges}

\begin{itemize}
\item[1.] Hot wind fluid. This component drives starburst-driven
  winds, yet has been detected in very few objects.  Metal abundances
  suggest enrichment by Type II supernovae, but the measurements are
  highly uncertain. Both sensitivity and high spatial resolution are
  needed to isolate the hot wind fluid from X-ray stellar binaries and
  the rest of the X-ray-emitting gas.  But, no such instrument is
  planned for the foreseeable future.
\item[2.] Entrained molecular gas and dust. Despite the important role
  of the molecular component in galactic winds, high-quality mm-wave
  data exist only for M82. This is due to the limited sensitivity and
  spatial resolution of current instruments, but this is changing. New
  mm-wave arrays (e.g., CARMA, and especially ALMA) will map the
  molecular gas in a large sample of nearby galaxies with excellent
  resolution ($<$ 1$^{\prime\prime}$). As described in \S 3.3.5, Spitzer
  has recently helped constrain the amount and location of dust in the
  winds.
\item[3.] Zone of influence and escape efficiency. Current estimates
  of these two quantities are limited by the sensitivity of the
  instrument used for the measurements. Deeper emission-line, X-ray,
  and radio maps of wind galaxies will provide better constraints on
  the extent of the wind and the probability that the outflowing
  material escapes from the potential well of the host.
\item[4.] Thermalization efficiency. Observational constraints on the
  thermalization efficiency of GWs are rare because of an incomplete
  accounting of the various sources of thermal energy and kinetic
  energy in the wind.  A multiwavelength approach that considers all
  gas phases is needed.
\item[5.] Wind / ISM interface and influence of magnetic fields.
  Constraints on microphysics at the interface between the wind and
  galaxy ISM are available in only a handful of galaxies.
  High-resolution ($\le$ parsec scale) imaging and spectra of the
  entrained disk material in a sizable sample of local objects are
  required. The large-scale morphology of the magnetic field lines has
  been mapped in a few winds, but the strength of the field on pc
  scale is unknown.  This information is crucial in estimating the
  conductivity between the hot and cold fluids.
\item[6.] Positive feedback. Star-forming radio jet/gas interactions
  have been found in a few nearby systems and are suspected to be
  responsible for the ``alignment effect'' between the radio and UV
  continua in distant radio galaxies. The same physics may also
  provide positive feedback in wind galaxies.  Convincing evidence for
  positive feedback has been found in the disk of M82 (Matsushida et
  al. 2004), but the frequency of this phenomenon is completely
  unknown.
\item[7.] Galactic winds in the distant universe. Absorption-line
  studies of high-$z$ galaxies and QSOs will remain a powerful tool to
  search for distant galactic winds and to constrain their
  environmental impact.  Future large ground and space telescopes will
  extend such studies to the reionization epoch.  These galaxies are
  very faint, but gravitational lensing by foreground clusters can
  make them detectable and even spatially resolved.
\end{itemize}

\section{Elemental Abundances as Tracers of Star Formation}

\subsection{Introduction: Basics of Chemical Evolution}
\label{}

Hydrogen, helium, and traces of lithium, boron, and beryllium were
produced early on in the Big Bang. All other elements ({\em i.e.} all
other ``metals'') were produced through nucleosynthesis in stars. The
abundances of these elements are therefore a direct tracer of past
star formation in a galaxy.

Gas is transformed into stars. Each star burns hydrogen and helium in
its nucleus and produces heavy elements. These elements are partially
returned into the interstellar gas at the end of the star's life via
stellar winds or supernovae explosions. Some fraction of the metals
are locked into the remnant of the star.  If there is no gas infall
from the outside or selective loss of metals to the outside, the metal
abundance of the gas, and of subsequent generations of stars, should
increase with time. So in principle the evolution of chemical element
abundances in a galaxy provides a clock for galactic aging. One should
expect a relation between metal abundances and stellar ages. On
average, younger stars should contain more iron than older stars.
This is partially the case for the solar neigborhood, where an
age-metallicity relation is seen for nearby disk stars, but a lot of
scatter is seen at old ages ($>$ 3 Gyr; e.g., Nordstrom, Andersen, \&
Mayor 2005). Clearly, our Galaxy is not as simple as described here
and we need to add a few more ingredients to better match the
observations.

In \S 4.2, I describe a few simple models to account for the complexity
of galaxies. An extensive literature exists on this topic. I refer the
readers to the seminal paper by Tinsley (1980) as well as Binney \&
Tremaine (1987; \S\S 9.2 and 9.3) and Binney \& Merrifield (1998; \S
5.3). In \S 4.3, I compare the predictions of these models with the
observations in local star-forming and starburst galaxies and in
distant star-forming galaxies and quasars. These comparisons help us
understand the integrated star formation history and chemical
evolution of these objects.

\subsection{Simple Models}
\label{}

All models discussed here assume that the galaxy's gas is well-mixed
{\em i.e.} uniform metal abundance, and that the (high-mass) stars
return their nucleosynthetic products rapidly, much faster than the
time to form a significant fraction of the stars (this is called the
``instantaneous recycling approximation'').

\subsubsection{Closed Box}

The closed-box model further assumes that no infall or outflow is
taking place. In that case, the total baryonic mass of the galaxy,
$M_{baryons} = M_{g(as)} + M_{s(tar)}$ = constant. If $Z$ is the
fraction by mass of heavy elements (the Sun's abundance is $Z_\odot
\sim 0.02$ and the most metal-poor stars in the Milky Way have $Z \le
10^{-4}$ $Z_\odot$), the mass of heavy elements in the gas $M_h = Z
M_g$.

If the total mass made into stars is $dM_s^\prime$ and the amount of
mass instantaneously returned to the ISM (from supernovae and stellar
winds, enriched with metals) is $dM_s^{\prime\prime}$, then the net
matter turned into stars is $dM_s = dM_s^\prime -
dM_s^{\prime\prime}$. The mass of heavy elements returned to the ISM
is $y~dM_s$, where $y$ is the yield of heavy elements (made
instantaneously). As a rule-of-thumb, only stars more massive than
$\sim$ 8 M$_\odot$ make heavies (supernovae). The fraction of mass
returned to the ISM $dM_s^{\prime\prime}/dM_s \sim 0.20$, the yield $y
\sim 0.01$ (dependent on stellar evolution and the Initial Mass
Function $\equiv$ IMF), and the metallicity of the shed gas $Z$(shed
gas) = (heavies shed) / (mass shed) = $y~dM_s / dM_s^{\prime\prime}$ $\sim$
0.01/0.2 = 0.05 ({\em i.e.} about 2.5 $\times$ $Z_\odot$).

In the closed-box model, mass conservation implies
\begin{equation}
dM_g + dM_s = 0 
\end{equation}
The net change in metal content of the gas is
\begin{equation}
dM_h = y dM_s - Z dM_s = (y - Z) dM_s 
\end{equation}
Since $dM_g = -dM_s$ and $Z = M_h / M_g$, the change in $Z$ is
\begin{eqnarray}
dZ & = & dM_h/M_g - M_h dM_g / M_g^2 \nonumber \\
   & = & (y - Z) dM_s / M_g + (M_h/M_g)(dM_s/M_g) = ydM_s / M_g \nonumber\\
dZ/dt & = & - y (dM_g/dt)/M_g \nonumber
\end{eqnarray}
Assuming $y$ = constant ({\em i.e.} independent of time and $Z$): 
\begin{eqnarray}
Z(t) & = & Z(0) - y~{\rm ln}~[M_g(t)/M_g(0)] \\
     & = & Z(0) - y~{\rm ln}~\mu(t), \nonumber
\end{eqnarray}
where $\mu$ = gas (mass) fraction $\equiv M_g(t)/M_g(0) = M_g(t)/M_t$.
The metallicity of the gas grows with time, as new stars are formed
and the gas is consumed.

The mass of the stars that have a metallicity less than $Z(t)$ is 
\begin{eqnarray}
M_s[< Z(t)] = M_s(t) = M_g(0) - M_g(t) = M_g(0)~[1 -e^{-(Z(t)-Z(0))/y}]
\end{eqnarray}
When all of the gas has been consumed, the mass of stars with
metallicity between $Z, Z + dZ$ is $dM_s(Z) \propto e^{-(Z-Z(0))/y}
dZ$. This exponential drop-off reproduces well the metallicity
distribution of stars in the bulge of our Galaxy (e.g., Rich 1990). 

The yield $y$ can be derived from observations:
\begin{eqnarray}
Z({\rm today}) \sim Z(0) - y~{\rm ln} [M(_g({\rm today})/M_g(0)]
\end{eqnarray}
The average metal content of the gas in the disk near the Sun is $Z
\sim 0.7 Z_\odot$. The initial mass of gas $M_g(0) = M_s({\rm today})
+ M_g({\rm today})$ where $M_s({\rm today}) \sim 40~M_\odot~{\rm
  pc}^{-2}$ and $M_g({\rm today}) \sim 10~M_\odot~{\rm pc}^{-2}$.
Assuming that $Z(0) = 0$, we derive $y \sim 0.43~Z_\odot$. Given this
value for the yield, we can compute the mass in stars with $Z < 0.25
Z_\odot$ compared to the mass in stars with the current metallicity of
the gas:
\begin{eqnarray}
M_s(< 0.25 Z_\odot) / M_s(< 0.7 Z_\odot) = [1 - e^{-0.25 Z_\odot /
  y}]/[1 - e^{-0.7 Z_\odot / y}] \sim 0.54
\end{eqnarray}
Therefore, half of all stars in the disk near the Sun should have $Z <
0.25~Z_\odot$. However, only 2\% of the F-G (old) dwarf stars in the
solar neighborhood have such metallicity. This discrepancy is known as
the ``{\em G-dwarf problem}''. Possible solutions to this problem
include (1) pre-enrichment in the gas: $Z(0) \sim 0.15 Z_\odot$, (2)
outflow (leaky-box model), and (3) infall (accreting-box
model). Solutions (2) and (3) are described next.

\subsubsection{Leaky Box}

If there is an outflow of processed material, $g(t)$, the conservation
of mass (Eq. 6) becomes:
\begin{eqnarray}
dM_g/dt + dM_s/dt + g(t) = 0
\end{eqnarray}
And the rate of change in the metal content of the gas mass (Eq. 7)
now becomes:
\begin{eqnarray}
dM_h/dt = y dM_s/dt - Z dM_s/dt - Z g(t) 
\end{eqnarray}
As a first-order approximation, one can assume that the rate at which
the gas flows out of the box is proportional to the star formation
rate: $g(t) = c~dM_s/dt$, where $c$ is a constant. In Lecture \#3 (\S
3.4), I noted that $c$ = 0.01 $\rightarrow$ 5 in starburst galaxies.
In that case, one finds that $dZ/dt = [y / M_g(t)] dM_s/dt$, where
$dM_s/dt = - (1/(1+c)) (dM_g/dt)$, so $dZ/dt = -[y/(1+c)] (1/M_g)
(dM_g/dt)$. Integrating this equation, we get
\begin{eqnarray}
Z(t) = Z(0) - [y / (1+c)] * {\rm ln} [M_g(t)/M_g(0)]
\end{eqnarray}
Comparing with Eq.\ (8), the only effect of an outflow is
therefore to reduce the yield to an {\em effective yield} = $[y /
(1+c)]$.

\subsubsection{Accreting Box}

Here we only consider the case of accretion of pristine (metal-free)
gas to the box. Since the gas is pristine, Eq. (7) is still valid:
the mass of heavy elements produced in a star formation episode is 
\begin{eqnarray}
dM_h/dt = (y - Z) dM_s/dt
\end{eqnarray}
However, Eq. (6) for the conservation of mass in the box becomes:
\begin{eqnarray}
dM_g/dt = -dM_s/dt + f(t),
\end{eqnarray}
where $f(t)$ is the accretion rate. Consider the simple case in which
the mass in gas in the box is constant. This implies then
\begin{eqnarray}
dZ/dt & = & (1/M_g) [(y - Z) dM_s/dt - Z dM_g/dt] \nonumber \\
      & = & (1/M_g) [(y - Z) dM_s/dt]
\end{eqnarray}
Integrating this equation and assuming that $Z(0) = 0$, 
\begin{eqnarray}
Z = y [1 - e^{-M_s/M_g}]
\end{eqnarray}
Therefore when $M_s >> M_g$, the metallicity $Z \sim y$. The mass in
stars that are more metal-poor than $Z$ is 
\begin{eqnarray}
M_s(< Z) = - M_g~{\rm ln}~(1 - Z/y)
\end{eqnarray}
In this case, for $M_g \sim 10~M_\odot~{\rm pc}^{-2}$ and $M_s \sim
40~M_\odot~{\rm pc}^{-2}$, and for $Z = 0.7~Z_\odot$, then $y \sim
0.71~Z_\odot$. Thus the fraction of stars more metal-poor than
0.25~$Z_\odot$ is $M(< 0.25)/M(< 0.7) \sim 10\%$, in much better
agreement with the observations of the solar neighborhood.

\subsection{Applications}
\label{}

It is technically easier to determine the chemical abundances in the
ISM than in stars so I only discuss the results on the gaseous
component of galaxies here. Elemental abundances in the ISM are
subject to major uncertainties so it is instructive to describe
briefly the general principles behind these measurements (interested
readers should refer to Edmunds \& Pagel 1984; McGaugh 1991; Kewley \&
Dopita 2002; Rupke, Veilleux, \& Baker 2008 for more detail). All
measurements rely on the relative strength of the various emission
lines produced by the gas heated and ionized by nearby hot, young
stars. The basic idea is to determine the abundance of each ionic
species and add them up to get the total abundance of a particular
element, e.g., for oxygen: O/H = O$^0$/H + O$^+$/H + O$^{++}$/H + ...
where H stands for the sum of neutral and ionized hydrogen.
Unfortunately, only a certain number of lines are strong enough to be
detected, so one often has to apply an ionization correction for the
species that are not directly observed.  These strong-line diagnostics
are calibrated against photoionization models, electron temperature
measurements ({\em i.e.} weak-line diagnostics), or a combination of
the two. Different line diagnostics, or different calibrations of the
same diagnostic, can give vastly different abundances for the same
galaxy or group of galaxies. So it is important when comparing
different galaxies to rely on the same line diagnostic with the same
calibration. The most commonly used strong-line diagnostics are
$R_{23} \equiv$ $\{f$([O~II] $\lambda\lambda$3726, 3729) + $f$([O~III]
$\lambda\lambda$4959, 5007)$\}$/$f$(H$\beta$) (Fig. 23) and $O_{32}
\equiv f$([O~III] $\lambda\lambda$4959, 5007)/$f$([O~II]
$\lambda\lambda$3726, 3729). The latter is a proxy for the ionization
parameter, which is the ratio of ionizing photons to hydrogen nuclei
present in the gas. The $f$([N~II] $\lambda$6583)/$f$([O~II]
$\lambda\lambda$3726, 3729) and $f$([N~II] $\lambda$6583)/$f$([S~II]
$\lambda\lambda$6716, 6731) ratios have also been used with some
success as metallicity indicators.

With these words of caution, I now proceed to describe the results of
recent analyses on local and distant galaxies.

\begin{figure}
\caption{ Comparison of the relation between metallicity and the line
  ratio [O~III] / [O~II]. The blue line shows the theoretical
  calibration of McGaugh (1991) for three representative values of 
  [O~III]/[O~II]. The green line shows the empirical calibration of
  Edmunds \& Pagel (1984), and the red line shows the semiempirical
  calibration of Zaritsky et al. (1994), itself the average of three
  previous calibrations. (From Tremonti et al. 2004)}
\end{figure}

\subsubsection{Local Star-Forming Galaxies}

\begin{figure}
\caption{Relation between stellar mass, in units of solar masses, and
  gas-phase oxygen abundance for 53,400 star-forming galaxies in the
  SDSS. The large black filled diamonds represent the median in bins
  of 0.1 dex in mass that include at least 100 data points. The solid
  lines are the contours that enclose 68\% and 95\% of the data. The 
  red line shows a polynomial fit to the data. The inset plot shows
  the residuals of the fit. (From Tremonti et al. 2004)}
  \end{figure}

  There is a well-known mass -- metallicity relation among local
  star-forming galaxies. Figure 24 shows the results from an analysis
  of 53,000 SDSS galaxies (Tremonti et al. 2004). As described in
  Lecture \#3 (\S 3.6.3) , this mass-metallicity relation is naturally
  explained by a leaky-box model which involves selective loss of
  metals via a galactic-scale outflow. Winds are more efficient at
  removing metals from shallower galaxy potential wells ($V_{rot} <
  150$ km s$^{-1}$; Garnett 2002) and the effective yield $y_{\rm eff}
  = [1/(1 + c)] y$ is smaller for smaller galaxies.

\begin{figure}
\caption{Oxygen abundance gradient in M101 from 20 H~II regions
  with electron temperature measurements. The linear fit to the data
  is shown by the solid line. (From Kennicutt et al. 2003)}
\end{figure}

The well-known metallicity -- radius relation {\em within} galaxies
also favors leaky-box models on a {\em local} scale. Figure 25 shows the
recent results from Kennicutt, Bresolin, \& Garnett (2003) on M~101
(see also Zaritsky, Kennicutt, \& Huchra 1994).

\subsubsection{Local Powerful Starburst Galaxies}

\begin{figure}
\caption{Ks-band luminosity-metallicity relation for nearby
  emission-line galaxies (small black circles), LIRGs (blue stars),
  and ULIRGs (red circles). Most of the LIRGs and ULIRGs fall well
  below the L-Z relation. The nearby galaxies are from the KISS
  sample, and the black line and dotted lines are a fit to the data
  and 1 $\sigma$ rms dispersion, respectively (Salzer et al. 2005).
  The dashed line locates solar abundance. The far-right points do not
  have measured K-magnitudes. The error bars represent the scatter in
  the abundance-R23 relation that was used to compute LIRG and ULIRG
  abundances, as well as the standard deviation in the ULIRG mass
  distribution. (From Rupke et al. 2008)}
\end{figure}

In a recent study of local luminous and ultraluminous infrared
galaxies (LIRGs and ULIRGs, respectively; see definitions in Lecture
\# 2; \S 2.1), Rupke et al.  (2008) have found that the oxygen
abundances (and effective yields) in the cores of these objects lie
significantly below the [O/H] -- host luminosity relation of local
star-forming galaxies (Fig. 26). They find that this effect increases
with increasing infrared luminosity (which is a proxy for the star
formation rate in the starburst; Fig. 27). They conclude that the
observed underabundance and smaller yield result from the combination
of a decrease of abundance with increasing radius in the progenitor
galaxies and strong, interaction- or merger-induced gas inflow into
the galaxy nucleus. This conclusion demonstrates that local abundance
scaling relations are not universal, a fact that must be accounted for
when interpreting abundances earlier in the universe's history, when
merger-induced star formation was the dominant mode (next section).

\begin{figure}
\caption{ Difference between the observed abundances in LIRGs and
  ULIRGs and the L-Z relation as a function of infrared luminosity.
  The small filled circles represent individual deviations from the
  L-Z relation. The thick open circles are median deviations from L-Z
  for equal-size bins centered on log[L$_{\rm IR}$/L$_\odot$] = 11.25,
  11.75, and 12.25 and the error bars represent the standard error in
  the mean in each bin. LIRGs are offset by 0.2 dex, and ULIRGs by 0.4
  dex. Comparison to the L-Z relation shows a mildly significant trend
  toward higher abundance offsets for higher $L_{\rm IR}$. (From Rupke
  et al.  2008)}
\end{figure}

\subsubsection{Distant Galaxies}

\begin{figure}
\caption{ Observed relation between stellar mass and oxygen abundance
  at $z \sim 2$, shown by the large red filled circles. Each point
  represents the average value of 14 or 15 galaxies, with the
  metallicity estimated from the [N II]/H$\alpha$ ratio of their
  composite spectrum.  Horizontal bars indicate the range of stellar
  masses in each bin, while the vertical error bars show the
  uncertainty in the [N~II]/H$\alpha$ ratio. The additional error bar
  in the lower right corner shows the additional uncertainty in the N2
  calibration itself. The dashed blue line is the best-fit
  mass-metallicity relation of Tremonti et al. (2004), shifted
  downward by 0.56 dex. The metallicities of different samples are
  best compared using the same calibration; we therefore show, with
  small gray dots, the metallicities of the 53,000 SDSS galaxies of
  Tremonti et al. (2004) determined with the N2 index. Note that the
  [N~II]/H$\alpha$ ratio saturates near solar metallicity (horizontal
  dotted line). The blue filled triangles indicate the mean
  metallicity of the SDSS galaxies in the same mass bins as the high-z
  sample; using the more reliable, low-metallicity bins, the high-z
  galaxies are 0.3 dex lower in metallicity at a given mass. (From Erb
  et al. 2006)}
\end{figure}

The mass-metallicity relation of distant galaxies appears to fall
below that of local galaxies. This is seen at 0.3 $< z <$ 1 (e.g.,
Kobulnicky \& Kewley 2004; Savaglio et al. 2005) and at higher
redshifts (e.g., Erb et al. 2006; Fig. 28). This result is usually
interpreted as being due to a redshift evolution -- in a closed box
model, the metallicity builds up with time. However, the results
discussed in \S 4.3.2 raise a red flag when interpreting the data at
high redshifts. It may be once again that lower abundances (and
effective yield) are due at least in part to merger-induced gas
inflows. Rupke et al. (2008) have tried to address this issue by
comparing galaxies with the same infrared luminosity at different
redshifts. They found that the oxygen abundance of LIRGs increases by
$\sim$ 0.2 dex from $z \sim 0.6$ to $z \sim 0.1$ (Fig. 29), while
modest if any evolution was found between $z \sim 2$ submm galaxies,
$z \sim 0.5$ ULIRGs and the local $z \sim$ ULIRGs.

\begin{figure}
\caption{Abundance offset from the local L-Z relation for low- and
  high-redshift LIRGs and ULIRGs, as a function of infrared
  luminosity. The black filled circles, blue open diamonds, and red
  open stars represent LIRGs and ULIRGs from Rupke et al. (2008),
  LIRGs and ULIRGs from Liang et al. (2004) and Rupke et al. (2008),
  and SMGs (Swinbank et al.  2004; Nesvadba et al. 2007),
  respectively. The black thick open circles, blue open diamonds, and
  open star are median deviations from L-Z for local LIRGs and ULIRGs,
  LIRGs, and SMGs, respectively.  The LIRGs clearly evolve upward in
  abundance by 0.2 dex from $z \sim 0.6$ to $z\sim 0.1$, as would be
  expected from continual processing of heavy elements.  Although
  there are only 2 ULIRGs in this figure, there is also apparent
  redshift evolution in ULIRG abundance from $z\sim 2$ to $z \sim
  0.1$. Finally, there is also evidence for modest evolution from SMGs
  to ULIRGs, although the observed scatter and systematic
  uncertainties are large. (From Rupke et al. 2008)}
\end{figure}

\subsubsection{Distant Quasars}

\begin{figure}
\caption{Measured NV/HeII and NV/CIV flux ratios versus redshift (left
  panels) and continuum luminosity (right). The upper and lower ranges
  might be undersampled (especially for NV/HeII at redshifts $>$ 1)
  because limits on weak lines (e.g. HeII) were often not available
  from the literature. The two asterisks in each panel represent mean
  values measured by Osmer, Porter, \& Green (1994) for high- and
  low-luminosity QSOs at redshift $>$3. The solid curves are
  predictions based on chemical evolution models. (From Hamann \&
  Ferland 1999) }
\end{figure}

Emission lines in distant quasars have been used successfully to
estimate the nuclear gas metallicity of the galaxy hosts. The most
luminous QSOs have nuclear metallicities typical of giant ellipticals.
A trend of increasing metallicity with increasing luminosity has been
found (e.g., Hamann \& Ferland 1999; Fig. 30). This may be equivalent
to the mass-metallicity measured at low redshift.  However,
interestingly, Shemmer et al. (2004) have found that metallicity is
even more strongly correlated with accretion rate onto the black hole
(which is a function of luminosity {\em and} H$\beta$ line width; Fig.
31). This may imply an intimate relation between starbursts,
responsible for the enrichment of the nuclear gas, and AGN fueling,
represented by the accretion rate.

\begin{figure}
\caption{N V/C IV vs. accretion rate. A strong metallicity-accretion
  rate correlation for all AGNs is apparent, and most NLS1s are found
  in the same region of parameter space that is shared by the high-z
  quasars. Note that the location of Mrk 766 (the NLS1 with the lowest
  accretion rate in this diagram) may be affected by strong intrinsic
  reddening (from Shemmer et al. 1994). }
\end{figure}


\begin{thebibliography}{}


\bibitem[Adelberger et al. (2005)]{ade05} Adelberger, K., et al. 2005,
  ApJ, 629, 636-653  
\bibitem[Alexander et al. (2005)]{ale05} Alexander, D., et al. 2005,
  ApJ, 632, 736-750         
\bibitem[Alexander et al. (2008)]{ale08} Alexander, D., et al. 2008,
  preprint. (astro-ph/0803.0634)
\bibitem[Bajtlik, Duncan, \& Ostriker (1988)]{baj88} Bajtlik, S.,
  Duncan, R. C., \& Ostriker, J. P. 1988, ApJ, 327, 570-583
\bibitem[Begelman (1985)]{beg85} Begelman, M. C. 1985, ApJ, 297, 492-506
\bibitem[Binney \& Merrifield (1998)]{bin98} Binney, J., \&
  Merrifield, M. 1998, Galactic Astronomy, Princeton, NJ: Princeton
  University Press, 796 p.
\bibitem[Binney \& Tremaine (1987)]{bin87} Binney, J., \& Tremaine,
  S. 1987, Galactic Dynamics, Princeton, NJ: Princeton University
  Press, 747 p.
\bibitem[Blain et al. (2002)]{bla02} Blain, A., et al. 2002, PhR, 369, 111-176
\bibitem[Bland-Hawthorn \& Cohen (2003)]{bla03} Bland-Hawthorn, J., \&
  Cohen,  2003. ApJ, 582, 246-256            
\bibitem[Barnes \& Hernquist (1996)]{bar96} Barnes, J. E., \&
  Hernquist, L. 1996, ApJ, 471, 115-142
\bibitem[Borys et al. (2005)]{bor05} Borys, C., et al. 2005, ApJ, 635,
  853-863  
\bibitem[Castor, McCray, \& Weaver (1975)]{cas75} Castor, J., McCray, R., \&
  Weaver, R. 1975, ApJ, 200, L107-L110
\bibitem[Cecil et al. (2001)]{cec01} Cecil, G., et al. 2001, ApJ, 555, 338-355
\bibitem[Cecil, Bland-Hawthorn, \& Veilleux (2002)]{cec02} Cecil, G.,
  Bland-Hawthorn, J., \& Veilleux, S. 2002, ApJ, 576, 745-752
\bibitem[Chapman et al. (2003)]{cha03} Chapman, S. C., et al. 2003,
  ApJ, 599, 92-104
\bibitem[Chapman et al. (2004)]{cha04} Chapman, S. C., et al. 2004,
  ApJ, 611, 732-738
\bibitem[Chapman et al. (2005)]{cha05} Chapman, S. C., et al. 2005,
  ApJ, 622, 772-796   
\bibitem[Cid Fernandes et al. (2004)]{cid04} Cid Fernandes, R.,  et
  al. 2004, MNRAS, 355, 273-296
\bibitem[Cid Fernandes \& Terlevich (1995)]{cid95} Cid Fernandes,
  R. Jr. \& Terlevich, R. 1995, MNRAS, 272, 423-441
\bibitem[Dasyra et al. (2006a)]{das06a} Dasyra, K., et al. 2006a, ApJ,
  638, 745-758            
\bibitem[Dasyra et al. (2006b)]{das06b} Dasyra, K.,  et al. 2006b,
  ApJ, 651, 835-852
\bibitem[Dasyra et al. (2007)]{das07} Dasyra, K., et al. 2007, ApJ,
  657, 102-115             
\bibitem[Devine \& Bally (1999)]{dev99} Devine, D., \& Bally J. 1999,
  ApJ, 510, 197-204
\bibitem[Downes \& Solomon (1998)]{dow98} Downes, D., \& Solomon,
  P. M. 1998, ApJ, 507, 615-654
\bibitem[Edmunds \& Pagel (1984)]{ed84} Edmunds, M. G., \& Pagel,
  B. E. J. 1984, MNRAS, 211, 507-519
\bibitem[Engelbracht et al. (2006)]{eng06} Engelbracht, C. W., et
  al. 2006, ApJ, 642, L127-L132
\bibitem[Erb et al. (2006)]{erb06} Erb, D. K., et al. 2006, ApJ, 644, 813-828
\bibitem[Evans et al. (2001)]{eva01} Evans, A. S., et al. 2001, AJ,
  121, 1893-1902
\bibitem[Ferrara (2007)]{fer07} Ferrara, A. 2007, EAS Publications
  Series, Vol. 24, pp. 229-243
\bibitem[Ferrarese \& Merrit (2000)]{fer00} Ferrarese, L., \& Merritt,
  D. 2000, ApJ, 539, L9-L12
\bibitem[Fukugita, Hogan, \& Peebles (1998)]{fuk98} Fukugita, M.,
  Hogan, C. J., \& Peebles, P. J. E. 1998, ApJ, 503, 518-530
\bibitem[Garnett (2002)]{gar02} Garnett, D. R. 2002, ApJ, 581, 1019-1031
\bibitem[Gebhardt et al. (2000)]{geb00} Gebhardt, K., et al. 2000,
  ApJ, 539, L13-L16
\bibitem[Genzel et al. (2001)]{gen01} Genzel, R., et al. 2001, ApJ,
  563, 527-545             
\bibitem[Gonz\'alez Delgado et al. (1998)]{gon98} Gonz\'alez Delgado,
  R. M., et al.  1998, ApJ, 495, 698-717
\bibitem[Gonz\'alez Delgado et al. (2001)]{gon01} Gonz\'alez Delgado,
  R. M., Heckman, T., \& Leitherer, C. 2001, ApJ, 546, 845-865
\bibitem[Griffiths et al. (2000)]{gri00} Griffiths, R. E., et
  al. 2000, Sci, 290, 1325-1328
\bibitem[Hamann \& Ferland (1999)]{ham99} Hamann, F., \& Ferland,
  G. 1999, ARA\&A, 37, 487-531
\bibitem[Heckman et al. (1997)]{hec97} Heckman, T. M., et al. 1997
  ApJ, 482, 114-132
\bibitem[Hoopes et al. (2005)]{hoo05} Hoopes, C. G., et al. 2005, ApJ,
  619, L99-L102
\bibitem[Hopkins et al. (2006)]{hop06} Hopkins, P. F., et al. 2006,
  ApJS, 163, 1-49
\bibitem[Ishida (2004)]{ish04} Ishida, K. 2004, PhD Thesis, University
  of Hawaii
\bibitem[Joguet et al. (2001)]{jog01} Joguet, B., A\&A, 380, 19-30
\bibitem[Kauffmann et al. (2003a)]{kau03} Kauffmann, G., et al. 2003a,
  MNRAS, 341, 54-69
\bibitem[Kauffman et al. (2003b)]{kau03} Kauffmann, G., et al. 2003b,
  MNRAS, 346, 1055-1077
\bibitem[Kennicutt, Bresolin, \& Garnett (2003)]{ken03} Kennicutt, R. C. Jr.,
  Bresolin, F., \& Garnett, D. R. 2003, apJ, 591, 801-820
\bibitem[Kennicutt (1998)]{ken98} Kennicutt, R. C. Jr. 1998, ARAA, 36, 189-231
\bibitem[Kewley et al. (2001)]{kew01} Kewley, L. J., et al. 2001, ApJ,
  556, 121-140
\bibitem[Kewley \& Dopita (2002)]{kew02} Kewley, L. J., \& Dopita,
  M. A. 2002, ApJS, 142, 35-52
\bibitem[Kobulnicky \& Kewley (2004)]{kob04} Kobulnicky, H. A., \&
  Kewley, L. J. 2004, ApJ, 617, 240-261
\bibitem[Kormendy \& Djorgovski (1989)]{kor89} Kormendy, J., \&
  Djorgovski, S. 1989, ARA\&A, 27, 235-277
\bibitem[Le Floc'h et al. (2005)]{lef05} Le Floc'h, E., et al. 2005,
  ApJ, 632, 169-190
\bibitem[Lehnert et al. (1999)]{leh99} Lehnert, M. et al. 1999, ApJ,
  523, 575-584
\bibitem[Liang et al. (2004)]{lia04} Liang, Y. C., et al. 2004, A\&A,
  423, 867-880
\bibitem[Lonsdale, Farrah, \& Smith (2006)]{lon06} Lonsdale, Farrah,
  \& Smith 2006, Astrophysics Update 2, edited by John W. Mason, p. 285
\bibitem[Lutz, Veilleux, \& Genzel (1999)]{lut99} Lutz, D., Veilleux,
  S., \& Genzel, R. 1999, ApJ, 517, L13-L17
\bibitem[Maiolino et al. (2007a)]{mai07a} Maiolino, R., et al. 2007a,
  A\&A, 468, 979-992
\bibitem[Maiolino et al. (2007b)]{mai07b} Maiolino, R., et al. 2007b,
  A\&A, 472, L33-L37
\bibitem[Martin, Kobulnicky, \& Heckman (2002)]{mar02} Martin, C.,
  Kobulnicky, H. A., \& Heckman, T. M. 2002, ApJ, 574, 663-692
\bibitem[Matsushita et al. (2004)]{mat04} Matsushita, S., et al. 2004,
  The Neutral ISM in Starburst Galaxies, ASP Vol. 320, eds. S. Aalto,
  S. Huttemeister, and A. Pedlar, p. 138
\bibitem[McGaugh (1991)]{mcg91} McGaugh, S. S. 1991, ApJ, 380, 140-150
\bibitem[Mihos \& Hernquist (1996)]{mih96} Mihos, J. C., \& Hernquist,
  L. 1996, ApJ, 464, 641-663
\bibitem[Mu\~noz Mar\'in et al. (2007)]{mun07} Mu\~noz Mar\'in, et
  al. 2007, AJ, 134, 648-667
\bibitem[Murray et al. (2005)]{mur05} Murray, N., Quataert, E., \& Thompson, T.
A. 2005, ApJ,  618, 569-585
\bibitem[Nesvadba et al. (2007)]{nes07} Nesvadba, N. P. H., et al. 2007,
  ApJ, 657, 725-737
\bibitem[Netzer et al. (2007)]{net07} Netzer, H., et al. 2007, ApJ,
  666, 806-816
\bibitem[Nordstrom, Andersen, \& Mayor (2005)]{nor05} Nordstrom,
  Andersen, \& Mayor 2005, in The Three-Dimensional Universe with
  Gaia, eds. C. Turon, K. S. O'Flaherty, M.A.C. Perryman, 183-186
\bibitem[Osmer et al. (1994)]{osm94} Osmer, P. S., Porter, A. C., \&
  Green, R. F. 1994, ApJ, 436, 678-695
\bibitem[Osterbrock et al. (1992)]{ost92} Osterbrock, D. E., Tran, H.,
  \& Veilleux, S. 1992, ApJ, 389, 196-207
\bibitem[Rich (1990)]{ric90} Rich, M. R. 1990, ApJ, 362, 604-619
\bibitem[Riffel et al. (2007)]{rif07} Riffer, R., et al. 2007, ApJ, 659, L103-L106
\bibitem[Robert et al. (1993)]{rob93} Robert, C., Leitherer, C., \&
  Heckman, T. M. 1993, ApJ, 418, 749-759
\bibitem[Rupke, Veilleux, \& Baker (2008)]{rup08} Rupke, D. S. R.,
  Veilleux, S., \& Baker, A. J. 2008, ApJ, 674, 172-193
\bibitem[Rupke, Veilleux \& Sanders (2005a)]{rup05a} Rupke, D. S. N.,
  Veilleux, S., \& Sanders, D. B. 2005a, ApJS, 160, 87-115
\bibitem[Rupke, Veilleux \& Sanders (2005b)]{rup05b} Rupke, D. S. N.,
  Veilleux, S., \& Sanders, D. B. 2005b, ApJS, 160, 115-148
\bibitem[Rupke, Veilleux \& Sanders (2005c)]{rup05c} Rupke, D. S. N.,
  Veilleux, S., \& Sanders, D. B. 2005c, ApJ, 632, 751-780
\bibitem[Salzer et al. (2005)]{sal05} Salzer, J. J., et al. 2005, ApJ,
  624, 661-679
\bibitem[Sanders \& Mirabel (1996)]{san96} Sanders, D. B., \& Mirabel,
  I. F. 1996, ARA\&A, 34, 749-792
\bibitem[Sanders et al. (1988a)]{san88a} Sanders, D. B., et al. 1988a, ApJ,
  325, 74-91
\bibitem[Sanders et al. (1988b)]{san88b} Sanders, D. B., et al. 1988b,
  ApJ, 328, L35-L39
\bibitem[Sanders et al. (1988c)]{san88c} Sanders, D. B., et al. 1988c,
  ApJ, 324, L55-L58
\bibitem[Savaglio et al. (2005)]{sav05} Savaglio, S.,  et al. 2005,
  ApJ, 635, 260-279
\bibitem[Schwartz \& Martin (2004)]{sch04} Schwartz, C. M., \& Martin,
  C. L. 2004, ApJ, 610, 201-212
\bibitem[Schweitzer et al. (2006)]{sch06} Schweitzer, M., et al. 2006,
  ApJ, 649, 79-90
\bibitem[Sedov (1959)]{sed59} Sedov, L. I. 1959, Similarity and
  Dimensional Methods in Mechanics, New York: Academic Press
\bibitem[Shapley et al. (2003)]{sha03} Shapley, A. E., et al. 2003,
  ApJ, 588, 65-89
\bibitem[Shemmer et al. (2004)]{she04} Shemmer, O., et al. 2004, ApJ,
  614, 547-5578
\bibitem[Shi et al. (2007)]{shi07} Shi, Y., et al. 2007, ApJ, 669,
  841-861
\bibitem[Shopbell \& Bland-Hawthorn (1998)]{sho98} Shopbell, P. L.,
  \& Bland-Hawthorn, J. 1998, ApJ, 493, 129-153
\bibitem[Silk (2003)]{sil03} Silk, J.  2003, MNRAS, 343, 249-254
\bibitem[Smail et al. (2002)]{sma02} Smail, I., et al. 2002, MNRAS,
  331, 495-520
\bibitem[Smail et al. (2004)]{sma04} Smail, I., et al. 2004, ApJ, 616,
  71-85
\bibitem[Sofue \& Handa (1984)]{sof84} Sofue, Y., \& Handa, T. 1984,
  Nature, 310, 568-569
\bibitem[Soifer et al. (2000)]{soi00} Soifer, B. T., et al. 2000, AJ,
  119, 509-523
\bibitem[Stevens, Read, \& Bravo-Guerrero (2003)]{ste03} Stevens, I. R., Read, A. M., \&  Bravo-Guerrero, J. 2003, MNRAS, 343, L47-L52
\bibitem[Swinbank et al.  (2004)]{swi04} Swinbank, A. M., et al.  2004,
  ApJ, 617, 64-80
\bibitem[Tacconi et al. (2002)]{tac02} Tacconi, L. J., et al. 2002,
  ApJ, 580, 73-87            
\bibitem[Taylor (1950)]{tay50} Taylor, G. 1950, Proc. R. Soc. London
  Ser. A 2001, 159
\bibitem[Terlevich et al. (1990)]{ter90} Terlevich, E., Diaz, A. I.,
  \& Terlevich, R. 1990, MNRAS, 242, 271-284
\bibitem[Tinsley (1980)]{tin80} Tinsley, B. M. 1980, Fund. of Cosmic
  Physics, 5, 287-388
\bibitem[Tran et al. (2001)]{tran01} Tran, Q. D., et al. 2001, ApJ,
  552, 527-543 
\bibitem[Tremonti et al. (2004)]{tre04} Tremonti, C. A., et al. 2004,
  ApJ, 613, 898-913
\bibitem[Veilleux (2001)]{vei01} Veilleux, S. 2001, in Starburst Galaxies: Near and Far, ed. L. Tacconi \& D. Lutz (Heidelberg: Springer), 88-95
\bibitem[Veilleux, Cecil, \& Bland-Hawthorn (2005)]{vei05} Veilleux,
  S., Cecil, G., \& Bland-Hawthorn, J. 2005, ARA\&A, 43, 769-826
\bibitem[Veilleux, Kim, \& Sanders (1999)]{vei99} Veilleux, S., Kim,
  D. C., \& Sanders, D. B. 1999, ApJ, 522, 113-138 
\bibitem[Veilleux, Kim, \& Sanders (2002)]{vei02} Veilleux, S., Kim,
  D. C., \& Sanders, D. B. 2002, ApJS, 143, 315-376
\bibitem[Veilleux \& Osterbrock (1987)]{vei87} Veilleux, S., \&
  Osterbrock, D. E. 1987, ApJS, 63, 295-310
\bibitem[Veilleux, Sanders, \& Kim (1999)]{vei99} Veilleux, S.,
  Sanders, D. B., \& Kim, D. C. 1999 , ApJ, 522, 139-156
\bibitem[Veilleux et al. (1994)]{vei94} Veilleux, S., et al. 1994,
  ApJ, 433, 48-64
\bibitem[Veilleux et al. (1995)]{vei95} Veilleux, S., et al. 1995,
  ApJS, 98, 171-217
\bibitem[Veilleux et al. (2003)]{vei03} Veilleux, S., et al. 2003,
  AJ, 126, 2185-2208
\bibitem[Veilleux et al. (2006)]{vei06} Veilleux, S., et al. 2006,
  ApJ, 643, 707-723          
\bibitem[Veilleux et al. (2008)]{vei08} Veilleux, S., et al. 2008, in prep. 
\bibitem[Walter, Weiss, \& Scoville (2002)]{wal02} Walter, F., Weiss,
  A., \& Scoville, N. 2002, ApJ, 580, L21-L25
\bibitem[Weaver et al. (1977)]{wea77} Weaver, R., et al. 1977, ApJ,
  218, 377-395
\bibitem[Wilson \& Tsvetanov (1994)]{wil94} Wilson, A. S., \&
  Tsvetanov, Z. 1994, AJ, 107, 1227-1234
\bibitem[Wisotzki (2000)]{wis00} Wisotzki, L. 2000, A\&A, 353, 853-860
\bibitem[Yuasa et al. (2007)]{yua07} Yuasa, T., et al. 2007, preprint,
  astro-ph/0709.1580
\bibitem[Zaritsky, Kennicutt, \& Huchra (1994)]{zar94} Zaritsky, D.,
  Kennicutt, R. C.  Jr., \& Huchra, J. P. 1994, ApJ, 420, 87-109

\end{thebibliography}
\end{document}